\documentclass[12pt,draftclsnofoot,onecolumn]{IEEEtran}

\usepackage{cite}
\usepackage{amsmath,amssymb,amsfonts}
\usepackage{algorithmic}
\usepackage[ruled,linesnumbered]{algorithm2e}
\usepackage{scalerel}

\usepackage{graphicx}
\usepackage{textcomp}
\usepackage{xcolor}

\usepackage{amsbsy}
\usepackage{gensymb}
\usepackage{enumitem}
\usepackage{amsthm}
\usepackage{amsmath,bm}
\usepackage{caption}
\usepackage{subcaption}

\theoremstyle{remark}

\def\BibTeX{{\rm B\kern-.05em{\sc i\kern-.025em b}\kern-.08em
    T\kern-.1667em\lower.7ex\hbox{E}\kern-.125emX}}
\begin{document}

\title{Transforming Fading Channel from Fast to Slow: Intelligent Refracting Surface Aided High-Mobility Communication
}

\author{Zixuan~Huang, \IEEEmembership{Student Member,~IEEE,} Beixiong~Zheng, \IEEEmembership{Member,~IEEE,} and~Rui~Zhang, \IEEEmembership{Fellow,~IEEE}

\thanks{Part of this work was presented in IEEE International Conference on Communications (ICC), Montreal, Canada, 2021 \cite{icc_this}.
}
\thanks{The authors are with the Department of Electrical and Computer Engineering, National University of Singapore, Singapore
117583 (e-mails: huang.zixuan@u.nus.edu, elezbe@nus.edu.sg, elezhang@nus.edu.sg). Z. Huang is also with the NUS Graduate School, National University of Singapore, Singapore 119077.}
}
\maketitle
\vspace*{-4em}
\begin{abstract}
Intelligent reflecting/refracting surface (IRS) has recently emerged as a promising solution to reconfigure wireless propagation environment for enhancing the communication performance by tuning passive signal reflection or refraction. In this paper, we study a new IRS-aided high-mobility communication system by employing the intelligent {\it refracting} surface with a high-speed vehicle to aid its passenger's communication with a remote base station (BS). Due to the environment's random scattering and vehicle's high mobility, a rapidly time-varying channel is typically resulted between the static BS and fast-moving IRS/user, which renders the channel estimation for IRS with a large number of passive refracting elements more challenging, as compared to that for the conventional slow fading IRS channels with low-mobility users. In order to reap the high IRS passive beamforming gain with low channel training overhead, we propose a new and efficient two-stage transmission protocol to achieve both IRS channel estimation and refraction optimization for data transmission. Specifically, by exploiting the quasi-static channel between the IRS and user both moving at the same high speed as well as the line-of-sight (LoS) dominant channel between the BS and IRS, the user first estimates the LoS component of the cascaded BS-IRS-user channel in Stage~I, based on which IRS passive refraction is designed to maximize the corresponding IRS-refracted channel gain. Then, the user estimates the resultant IRS-refracted channel as well as the non-IRS-refracted channel in Stage~II for setting an additional common phase shift at all IRS refracting elements so as to align these two channels for maximizing the overall channel gain for data transmission. Simulation results show that the proposed design can efficiently achieve the full IRS passive beamforming gain in the high-mobility communication scenario, which also converts the overall BS-user channel from fast to slow fading for more reliable transmission. The proposed on-vehicle IRS system is further compared with a baseline scheme of deploying fixed IRSs (intelligent reflecting surfaces) on the roadside to assist high-speed vehicular communications, which achieves significant rate improvement due to its greatly saved channel training time.
\end{abstract}


\section{Introduction}
In the last few decades, the demands for increasingly higher data rates, more reliable network coverage and connectivity, as well as lower latency have driven significant advances in wireless communications. To meet these demands, various wireless technologies have been studied and developed, such as full-duplex (FD) relaying, small-cell base station (BS), distributed antennas/remote radio heads (RRHs), massive multiple-input multiple-output (MIMO), millimeter wave (mmWave) communications \cite{aa1,aa2,aa3}, etc.
However, these technologies will face critical challenges in future wireless systems such as 6G due to the higher operating frequencies that can incur substantially increased costs in hardware, power consumption, and signal processing \cite{en2}. Moreover, the random and time-varying wireless channels are becoming the ultimate bottleneck in achieving the targeted ultra-reliable and low-latency wireless communication (URLLC). Existing approaches to overcome this bottleneck either adapt to the wireless channel fading with dynamic resource allocation and beamforming designs, or compensate for the channel deep fading via advanced diversity, modulation and coding techniques \cite{gold,tse}, whereas the wireless propagation environment still remains largely uncontrollable.  

Recently, intelligent reflecting/refracting surface (IRS) and its various equivalents have emerged as a cost-effective solution to achieve smart and reconfigurable radio environment via tunable signal reflection or refraction \cite{qq1,tutorial,tut_2}. Specifically, IRS is a digitally-controllable metasurface consisting of a massive number of passive reflecting/refracting elements \cite{tunable1,tunable2,tunable3}, whose amplitudes and/or phase shifts can be individually  controlled in real time, thereby enabling  dynamic control over the wireless propagation channel for a variety of purposes (e.g., passive relaying/beamforming and interference nulling/cancellation \cite{tutorial}). Moreover, IRS dispenses with radio frequency (RF) chains and only reflects/refracts the ambient signals passively, which thus features low hardware cost and energy consumption. As such, IRS has been extensively investigated for various wireless systems and applications such as MIMO \cite{mimo1,mimo2}, orthogonal frequency division multiplexing (OFDM) \cite{ofdmi1,ofdmb1,ofdmb2,ofdmnew}, non-orthogonal multiple access (NOMA) \cite{normabei,norma1}, cognitive radio \cite{cr1,cr2}, simultaneous wireless information and power transfer (SWIPT) network \cite{swipt1,swipt2,pch_s1,pch_s2}, secrecy communication \cite{sec1,sec2},
and so on. 
More details on IRS-aided wireless communications for different systems and applications can be found in e.g., \cite{tutorial,tut_emi,tut_1,tut_2,tut_3}.

\begin{figure}[t]
\centering
\includegraphics[width=0.7\textwidth]{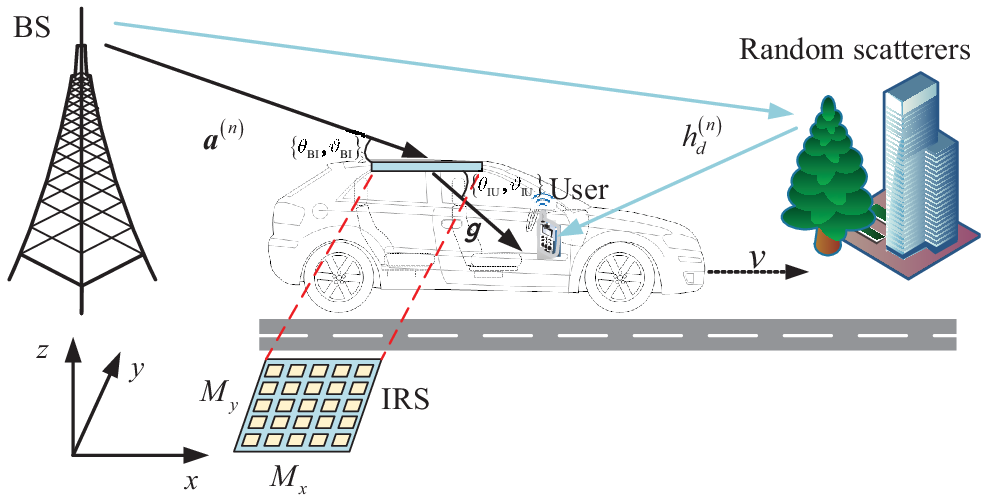}
\caption{Proposed on-vehicle IRS-aided communication system for high-mobility user.}
\label{config}
\vspace*{-2em}
\end{figure}

To reap the high passive beamforming gain of IRS, the acquisition of accurate channel state information (CSI) for the links between the IRS and its associated BS/users is crucial \cite{bsurv}, which, however, is practically challenging due to the following reasons. First, IRS can perform passive signal reflection/refraction only, which renders the CSI acquisition by conventional approaches requiring active transmitter/receiver infeasible \cite{qq1,tutorial}. As such, an alternative approach in practice is to estimate the cascaded BS-IRS-user channels based on the pilot symbols sent by the users/BS with properly designed IRS reflection/refraction patterns over time (see, e.g., \cite{ce1,ofdmi1}). Second, IRS generally consists of a vast number of passive elements, which can incur prohibitively high training overhead for channel estimation and thus severely degrade the throughput for data transmission. In order to reduce the training overhead, an effective method is to group adjacent reflecting/refracting elements with high spatial channel correlation into a subsurface, and thus only the effective cascaded BS-IRS-user channel associated with each sub-surface (instead of those for its individual elements) need to be estimated \cite{ofdmi1,partichang}. In \cite{tq_rand}, the authors considered random beamforming to avoid the high training overhead. In \cite{yxj_1}, the authors proposed an algorithm to reduce the training overhead via sparse matrix factorization and completion. In \cite{liuliang2}, efficient channel estimation for IRS-assisted multiuser communications was achieved by exploiting the common BS-IRS channel. In \cite{hxl,pch_statsical}, location information/statsitical CSI was exploited to reduce the training overhead. By exploiting the prior knowledge of the slow-varying LoS dominant BS-IRS channel, the authors in \cite{liuhang1} proposed an efficient algorithm for cascaded channel estimation.

However, existing works on IRS have mostly considered its reflecting mode for assisting the communications of low-mobility users (i.e., slow fading channels to/from the IRS) with one or more IRSs deployed at fixed locations, which are generally inapplicable to high-mobility scenarios such as high-speed vehicular communication. In such case, due to the environment's random scattering and vehicle's high speed, the transmitted signal from the BS usually arrives at the user's receiver over multiple propagation paths with rapidly time-varying phase shifts at different Doppler frequencies, which leads to a superimposed fast fading channel (i.e., the overall channel's amplitude and phase both vary substantially over time) \cite{gold}. As a result, the communication performance in terms of achievable rate or non-outage probability can be severely degraded. In \cite{bass}, IRSs deployed at fixed locations on the roadside were proposed to assist the high-speed vehicular communication by tuning the IRS reflection to compensate for the severe Doppler effect. In \cite{dsce}, IRS channel estimation was investigated with the Doppler effect taken into account. 
However, these studies assumed that each IRS is deployed at a fixed location, which renders very limited time for it to assist the communication between the BS and a high-mobility user passing by it. 

To circumvent the above difficulty, we consider a new IRS-aided high-mobility communication system in this paper, where IRS with a large number of {\it refracting} elements is employed with a high-speed vehicle (such as car, train, etc.) to aid the communication between the user (passenger) residing in it and a remote BS, as shown in Fig.~\ref{config}.
To reap the full passive beamforming gain of the IRS with its optimally tuned passive refraction, the acquisition of the accurate CSI for each refracting element/sub-surface is crucial. However, this problem is more practically challenging in our considered IRS-aided high mobility communication scenario as compared to the conventional low-mobility cases with slow fading channels to/from the IRS, which were considered in most of the existing literature on IRS.  
Specifically, due to user's high mobility, the IRS-refracted path and the other non-IRS-refracted paths all change in phase more rapidly over time, which in general requires more frequent channel training/estimation and thus can reduce the data transmission time substantially.
To tackle this challenge in IRS-aided high mobility communication, we propose a new and customized transmission protocol to conduct efficient channel estimation and refraction design for the proposed on-vehicle IRS system to achieve high communication rate and yet low outage probability. 
The main contributions of this paper are summarized as follows:
\begin{itemize}
\item First, our proposed new transmission protocol for the on-vehicle IRS exploits both the quasi-static channel between the IRS and user both moving at the same high speed (as the vehicle) and the line-of-sight (LoS) dominant channel between the BS and IRS. 
Specifically, the user first estimates the LoS component of the cascaded BS-IRS-user channel, based on which the IRS passive refraction is designed to maximize the LoS path gain of the IRS-refracted channel for data transmission. 
Subsequently, the user estimates the effective IRS-refracted channel and the non-IRS-refracted channel, based on which an additional common phase shift is set for all IRS refracting elements to align these two channels for maximizing the received signal power at the user for data transmission, which also effectively converts the overall BS-user channel from fast to slow fading to reduce the outage probability. 

\item Next, we present the detailed algorithm for estimating the essential parameters that characterize the LoS component of the IRS-refracted channel. 
However, this problem is a non-convex optimization problem, which is thus difficult to be optimally solved.
As such, we propose an efficient two-step algorithm to estimate the required channel parameters sub-optimally. 
Specifically, to reduce the computational complexity, we first perform a coarse two-dimensional (2D) grid-based search to find an initial estimate. 
Then, we refine the estimate by applying a gradient-based algorithm.

\item Last, we provide extensive simulation results to evaluate the performance of the proposed system and design. 
We show that our proposed design is effective in converting the end-to-end BS-user channel from fast to slow fading and also achieves significant rate improvement over the conventional design for slow-fading IRS channels with low-mobility users, for which the full CSI (of both the IRS-refracted and non-IRS-refracted channels) need to be estimated during each (short) channel coherence interval under the high-mobility communication scenario. 
Moreover, we demonstrate that the proposed vehicle-side IRS (Intelligent Refracting Surface) system is more efficient in enhancing the user's communication performance in a high-speed vehicle, as compared to a baseline roadside IRS (Intelligent Reflecting Surface) system that requires close-by IRSs deployed with fixed intervals on the roadside to aid the high-mobility users passing by.
\end{itemize}

The rest of this paper is organized as follows. 
Section~II presents the system model for the proposed vehicle-side IRS-aided high-mobility communication system. 
In Section~III, we propose a two-stage transmission protocol for channel estimation and IRS refraction design. 
Simulation results and discussions are presented in Section IV. 
Finally, conclusions are drawn in Section V.


{\emph{Notations:}} 
Upper-case and lower-case boldface letters denote matrices and column vectors, respectively. Upper-case calligraphic letters (e.g., $\mathcal{T}$) denote discrete and finite sets. Superscripts $(\cdot)^{T}$, $(\cdot)^{*}$, $(\cdot)^{H}$, and $(\cdot)^\dagger$ stand for the transpose, conjugate, Hermitian transpose, and Moore-Penrose inverse operations, respectively.
$x \operatorname{mod} y$ denotes $x$ modulo $y$.
$\left\|\cdot\right\|$ denotes the $l_2$ norm.
$\left\lfloor \cdot \right\rfloor$ denotes the floor operation.
$\angle (\cdot)$ denotes the angle of a complex number.
$\mathbb{R}^{x \times y}$ denotes the space of $x \times y$ real matrices.
$\mathbb{C}^{x \times y}$ denotes the space of $x \times y$ complex matrices.
$[\cdot]_{i,j}$ denotes the $(i,j)$-th element of a matrix.
$\nabla f (\boldsymbol{x})$ denotes the gradient of a scalar function $f (\boldsymbol{x})$.
$\operatorname{diag}(\boldsymbol{x})$ denotes a square diagonal matrix with the elements of $\boldsymbol{x}$ on the main diagonal.
$\otimes$ denotes the Kronecker product. 
$\odot$ denotes the Hadamard product. 
$\boldsymbol{\mathrm{I}}_{M}$ denotes an identity matrix with its dimension of $M$.
$\boldsymbol{1}_{M}$ denotes an all-one vector with its dimension of $M$.
$\mathrm{Re}\{\cdot\}$ and $\mathrm{Im}\{\cdot\}$ denote the real and imaginary parts of a complex number, respectively.
$\mathcal{O}(\cdot)$ denotes the big O notation.
$\mathbb{E}\{\cdot\}$ denotes the statistical expectation.
The distribution of a circularly symmetric complex Gaussian (CSCG) random variable with mean $\mu$ and variance $\sigma^2$ is denoted by $\mathcal{N}_{c}\left(\mu, \sigma^{2}\right)$; and $\sim$ stands for “distributed as”.

\section{System Model}
As shown in Fig.~\ref{config}, we consider an IRS-aided high-mobility communication system in the rural area, where an IRS is deployed at the top/side of a high-speed vehicle (to replace the metal panel that incurs high penetration loss) to aid its passenger/user's communication with a static BS. 
We assume that the vehicle moves at a high speed of $v$ meters/second (m/s). 
For the purpose of exposition, we consider the downlink communication and assume that both the BS and user\footnote{For the general multi-user case, the quasi-static locations of different users in a high-speed vehicle can be exploited to change the IRS refraction to serve them in a time division multiple access (TDMA) manner.} terminal are equipped with a single antenna\footnote{The results of this paper are also applicable to the uplink communication as well as the case of multi-antenna BS if the precoding vector at the BS has been set separately and fixed.}. 
Moreover, we assume that the IRS is a uniform planar array (UPA) composed of $M=M_x \times M_y$ refracting elements placed in the $x-y$ plane in the three-dimensional (3D) Cartesian coordinate system (see Fig.~\ref{config}), which is connected to a smart controller that is able to adjust its refracting elements’ individual on/off status and phase shifts, and also exchange (control/channel) information with the BS/user via separate reliable wireless links.
In this paper, we focus on one typical transmission frame of duration $T$, which is divided into $N$ time blocks (denoted by the set $\mathcal{N} \triangleq\left\{1, \ldots, N\right\}$), each with an equal duration of $T_b = T/N$.

Let $h^{\left(n\right)}_\mathrm{d}$ denote the baseband equivalent channel for the direct link from the BS to the user without any IRS refraction during  block $n \in \mathcal{N}$, which is assumed to remain constant during each block, but may change from one block to another due to the vehicle/user's mobility.
Let $\boldsymbol{s}\left( \phi,\bar{M} \right) = \left[1,e^{j \pi \phi}, \ldots, e^{j( \bar{M}-1 )\pi \phi}\right]^T$ denote the one-dimensional (1D) steering vector function of the IRS (applicable to both of its signal receiving and refracting), where $\phi$ denotes the phase difference (normalized to $\pi$) between any two adjacent elements and $\bar{M}$ denotes the number of elements involved.
Considering that the IRS is mounted at the top of the vehicle for example, we assume that the BS-IRS channel is LoS-dominant, which is denoted by $\boldsymbol{a}^{\left(n\right)} \in \mathbb{C}^{M \times 1}$. 
Note that as the IRS moves along with the high-speed vehicle, the LoS component of the BS-IRS channel mainly experiences the Doppler-induced phase shifts over different blocks, while its non-LoS (NLoS) component usually varies over blocks in both amplitude and phase due to the environment's random and multi-path scattering. 

Thus, we consider the Rician fading channel for the BS-IRS link, which is modelled as
\begin{equation}\label{bs-irs}
\boldsymbol{a}^{\left(n\right)} = 
\underbrace{\sqrt{\frac{ K}{1+K}} \rho e^{2\pi f_d (n-1) T_b}}_{\alpha_\mathrm{BI}^{\left(n\right)} }
\underbrace{\boldsymbol{s}\left( \phi_\mathrm{BI},M_x \right) \otimes \boldsymbol{s}\left( \varphi_\mathrm{BI},M_y \right)}_{\boldsymbol{s}_\mathrm{BI}\left( \theta_\mathrm{BI},\vartheta_\mathrm{BI} \right)}
+  \boldsymbol{a}_\mathrm{NLoS}^{\left(n\right)}, \quad \forall n \in \mathcal{N},
\end{equation} 
where $\alpha_\mathrm{BI}^{\left(n\right)} \in \mathbb{C}$ denotes the complex-valued channel gain of the LoS component during block $n$ by taking into account the Doppler effect\footnote{We assume that the Doppler-induced phase shift remains approximately constant within each block $n$ due to its short duration.}, with $\rho\in \mathbb{C}$ being a constant, $f_d = v \cos \theta_\mathrm{BI} \cos \vartheta_\mathrm{BI} /\lambda$ being the Doppler frequency, $\lambda$ being the signal wavelength,  $\theta_\mathrm{BI} \in \left[0,\pi /2\right]$ and $\vartheta_\mathrm{BI} \in \left[0,2\pi\right)$ being the elevation and azimuth angles-of-arrival (AoAs) at the IRS, respectively, $K$ being the Rician factor; 
$\boldsymbol{s}_\mathrm{BI}\left( \theta_\mathrm{BI},\vartheta_\mathrm{BI} \right)$ denotes the receive array response vector of the IRS, with $\phi_\mathrm{BI} \triangleq \frac{2 d}{\lambda} \cos \theta_\mathrm{BI} \cos \vartheta_\mathrm{BI} \in \left[-\frac{2 d}{\lambda}, \frac{2 d}{\lambda}\right]$ and $\varphi_\mathrm{BI} \triangleq \frac{2 d}{\lambda}  \cos \theta_\mathrm{BI} \sin \vartheta_\mathrm{BI} \in \left[-\frac{2 d}{\lambda}, \frac{2 d}{\lambda}\right]$, $d$ denoting the spacing between any two adjacent IRS elements along the $x$ or $y$-dimension, and $\boldsymbol{a}_\mathrm{NLoS}^{\left(n\right)} \sim \mathcal{N}_{c}\left(\boldsymbol{0},\frac{\left|\rho\right|^2}{1+K}  \boldsymbol{\mathrm{I}}_{M} \right)$ denotes the NLoS (modelled as Rayleigh fading) component during block $n$.
Note that for the BS-IRS channel in (\ref{bs-irs}), we assume the parameters $\rho$, $f_d$, $\theta_\mathrm{BI}$, and $\vartheta_\mathrm{BI}$ remain approximately constant within one transmission frame of interest. 
This assumption is practically valid due to the following reasons.
First, the traveling distance of the vehicle within one transmission frame is negligible as compared to the nominal distance with the remote BS, thus resulting in only marginal variations in the geometry-related parameters, i.e., $\{\rho,\theta_\mathrm{BI},\vartheta_\mathrm{BI}\}$ \cite{gao1}. Second, the vehicle speed $v$ is nearly constant during one transmission frame, which leads to the constant $f_d$.

On the other hand, due to the short distance between the IRS and user as well as the fact that they remain relatively static (despite that they both move at a high speed of the vehicle), the IRS-user channel changes much more slowly as compared to the BS-IRS channel. 
Thus, it is practically quasi-static and can be assumed to be an approximately constant LoS channel during each transmission frame, which is given by   
\begin{equation}\label{irs-user}
\boldsymbol{g} = \alpha_\mathrm{IU} 
\underbrace{\boldsymbol{s}\left( \phi_\mathrm{IU},M_x \right) \otimes \boldsymbol{s}\left( \varphi_\mathrm{IU},M_y \right)}_{\boldsymbol{s}_\mathrm{IU}\left( \theta_\mathrm{IU},\vartheta_\mathrm{IU} \right)},
\end{equation} 
where $\alpha_\mathrm{IU}\in \mathbb{C}$ denotes the complex-valued channel gain of the IRS-user link, $\theta_\mathrm{IU} \in \left[-\pi/2,0\right]$ and $\vartheta_\mathrm{IU} \in \left[0,2\pi\right)$ denote the elevation and azimuth angles-of-departure (AoDs) from the IRS to the user, respectively, and $\boldsymbol{s}_\mathrm{IU}\left(\theta_\mathrm{IU}, \vartheta_\mathrm{IU}\right)$ represents the refraction array response vector of the IRS with $\phi_\mathrm{IU} \triangleq \frac{2 d}{\lambda} \cos \theta_\mathrm{IU} \cos \vartheta_\mathrm{IU} \in \left[-\frac{2 d}{\lambda}, \frac{2 d}{\lambda}\right]$ and $\varphi_\mathrm{IU} \triangleq \frac{2 d}{\lambda} \cos \theta_\mathrm{IU} \sin \vartheta_\mathrm{IU} \in \left[-\frac{2 d}{\lambda}, \frac{2 d}{\lambda}\right]$.

Let $\boldsymbol{\nu}^{\left(n\right)} = \eta \left[ e^{j \omega^{\left(n\right)}_{1}},  \ldots,e^{j  \omega^{\left(n\right)}_{M}}\right]^T \in \mathbb{C}^{M \times 1 }$ denote the IRS refraction vector during block $n$, where $\eta$ denotes the IRS refraction amplitude, which is usually smaller than one due to practical penetration loss, and $\omega^{\left(n\right)}_{m}$ denotes the refraction phase shift of element $m$. 
The cascaded BS-IRS-user channel with the IRS refraction (or IRS-refracted channel) taken into account is thus expressed as
\begin{align}\label{hr}
h^{\left(n\right)}_\mathrm{r}
= \left( \boldsymbol{\nu}^{\left(n\right)} \right)^T
\underbrace{\left(\boldsymbol{g}\odot \boldsymbol{a}^{\left(n\right)} \right)}_{\boldsymbol{c}^{\left(n\right)}}
= \left( \boldsymbol{\nu}^{\left(n\right)} \right)^T
\left(\boldsymbol{c}_\mathrm{LoS}^{\left(n\right)} +\boldsymbol{c}_\mathrm{NLoS}^{\left(n\right)} \right) 
, \quad \forall n \in \mathcal{N},  
\end{align}
where $\boldsymbol{c}^{\left(n\right)} \triangleq \boldsymbol{c}_\mathrm{LoS}^{\left(n\right)} +\boldsymbol{c}_\mathrm{NLoS}^{\left(n\right)} \in \mathbb{C}^{M \times 1 }$ denotes the BS-IRS-user cascaded channel without considering IRS phase shifts, with i): $\boldsymbol{c}_\mathrm{LoS}^{\left(n\right)}$ denoting the LoS component, which is further expressed as 
\begin{align}\label{stee_los}
\boldsymbol{c}_\mathrm{LoS}^{\left(n\right)} 
&=  \alpha_\mathrm{BI}^{\left(n\right)}  \boldsymbol{g}\odot \boldsymbol{s}_\mathrm{BI}\left(\theta_\mathrm{BI},\vartheta_\mathrm{BI}\right)
\nonumber \\
&= \underbrace{ \alpha_\mathrm{BI}^{\left(n\right)}  \alpha_\mathrm{IU}}_{\beta^{\left(n\right)}}
\boldsymbol{s}_\mathrm{IU}\left(\theta_\mathrm{IU}, \vartheta_\mathrm{IU}\right) \odot \boldsymbol{s}_\mathrm{BI}\left(\theta_\mathrm{BI}, \vartheta_\mathrm{BI}\right)
\nonumber \\
&\stackrel{(a)}{=}\beta^{\left(n\right)}
\underbrace{
\left(\boldsymbol{s}\left( \phi_\mathrm{IU}, M_x \right) \odot \boldsymbol{s}\left( \phi_\mathrm{BI}, M_x \right)  \right) }_{\boldsymbol{s} \left( \tilde{\psi}_x, M_x \right)}
\otimes 
\underbrace{
\left(\boldsymbol{s}\left( \varphi_\mathrm{IU}, M_y \right) \odot \boldsymbol{s}\left( \varphi_\mathrm{BI}, M_y \right)  \right) }_{\boldsymbol{s} \left( \tilde{\psi}_y, M_y \right)}
\nonumber\\
&=\beta^{\left(n\right)} \boldsymbol{u}\left( \tilde{\psi}_x, \tilde{\psi}_y \right)
, \quad \forall n \in \mathcal{N},
\end{align}
where $\beta^{\left(n\right)}\in \mathbb{C}$ denotes the product channel gain of the LoS component, $(a)$ is obtained according to the mixed-product property of Kronecker product,  $\tilde{\psi}_x=\phi_\mathrm{IU}+\phi_\mathrm{BI} \in \left[-\frac{4 d}{\lambda}, \frac{4 d}{\lambda}\right]$, $\tilde{\psi}_y=\varphi_\mathrm{IU}+\varphi_\mathrm{BI} \in \left[-\frac{4 d}{\lambda}, \frac{4 d}{\lambda}\right]$, and $\boldsymbol{u}\left( \tilde{\psi}_x, \tilde{\psi}_y \right) = \boldsymbol{s} \left( \tilde{\psi}_x, M_x \right) \otimes \boldsymbol{s} \left( \tilde{\psi}_y, M_y \right)$ represents the 2D steering vector of the IRS; and ii):
\begin{align}\label{NLOS_RE}
\boldsymbol{c}_\mathrm{NLoS}^{\left(n\right)}
=  \alpha_\mathrm{IU}  \boldsymbol{s}_\mathrm{IU}\left(\phi_\mathrm{IU},\varphi_\mathrm{IU}\right) \odot \boldsymbol{a}_\mathrm{NLoS}^{\left(n\right)} 
,  \quad \forall n \in \mathcal{N}, 
\end{align} 
denoting the NLoS component. 
As $\boldsymbol{s}\left( \phi, \bar{M} \right)$ is a periodic function of $\phi$ with period 2,
we define $\psi_x \triangleq \tilde{\psi}_x \operatorname{mod}  2 \in \left[-1, 1\right]$ $\left(\psi_y \triangleq \tilde{\psi}_y \operatorname{mod}  2 \in \left[-1, 1\right]\right)$ as the cascaded effective phase along the $x$-axis ($y$-axis), 
such that we have $\boldsymbol{u}\left( \tilde{\psi}_x, \tilde{\psi}_y \right) = \boldsymbol{u}\left( \psi_x, \psi_y \right)$.
Hence, the end-to-end channel between the BS and user by combining the BS-user direct channel and the BS-IRS-user cascaded channel is given by 
\begin{align}
h^{\left(n\right)} 
&= \left( \boldsymbol{\nu}^{\left(n\right)} \right)^T
\underbrace{\left(\boldsymbol{c}_\mathrm{LoS}^{\left(n\right)} +\boldsymbol{c}_\mathrm{NLoS}^{\left(n\right)} \right)}_{\boldsymbol{c}^{\left(n\right)}} + h^{\left(n\right)}_\mathrm{d}, \quad \forall n \in \mathcal{N}.
\label{equiva}
\end{align}

Based on the channel model in (\ref{equiva}) and considering the ideal case where the CSI of $\boldsymbol{c}^{\left(n\right)}$ and $h^{\left(n\right)}_\mathrm{d}$ is perfectly available, the optimal IRS refraction vector $\boldsymbol{\nu}^{\left(n\right)}_\mathrm{opt}$ for block $n$ that maximizes the power of the end-to-end channel $h^{\left(n\right)}$ for data transmission is given by \cite{qqz}
\begin{align}
\left[\boldsymbol{\nu}^{\left(n\right)}_\mathrm{opt}\right]_m
&= e^{j \left(-\angle \left[\boldsymbol{c}^{\left(n\right)}\right]_m +\angle h^{\left(n\right)}_\mathrm{d}\right)}, \quad \forall m \in \{1, \ldots, M \}, \quad \forall n \in \mathcal{N}.
\label{optimal_vec}
\end{align}
However, the acquisition of such full CSI of $\boldsymbol{c}^{\left(n\right)}$ and $h^{\left(n\right)}_\mathrm{d}$ for each block $n$ may require a prohibitively high training overhead as that for $\boldsymbol{c}^{\left(n\right)}$ is in general proportional to the number of refracting elements $M$ \cite{ofdmb1,partichang}, which is usually very large in practice. 
As a result, considering the short interval of each block in the typical high-mobility communication scenario, the time for data transmission in each block will be severely reduced, which may even overwhelm the IRS beamforming gain and thus result in even lower communication throughput as compared to that of the conventional system without IRS. 
Nevertheless, it is worth noting that in our proposed system, the IRS is deployed with the high-speed vehicle, thus an LoS-dominant channel is very likely to occur for the BS-IRS link, which also leads to an LoS-dominant BS-IRS-user cascaded channel. 
Motivated by this, instead of estimating $\boldsymbol{c}^{\left(n\right)}$ exactly, we propose to estimate its LoS component $\boldsymbol{c}_\mathrm{LoS}^{\left(n\right)}$ for reducing the channel training overhead yet without sacrificing the gain of IRS passive beamforming designed based on the estimated $\boldsymbol{c}_\mathrm{LoS}^{\left(n\right)}$.
In particular, although the path gain $\beta^{\left(n\right)}$ of $\boldsymbol{c}_\mathrm{LoS}^{\left(n\right)}$ is time-varying over different blocks due to the Doppler effect, the effective phases $\{\psi_x,\psi_y\}$ in $\boldsymbol{c}_\mathrm{LoS}^{\left(n\right)}$ remain approximately unchanged within the entire transmission frame, which can be exploited for significantly reducing the training overhead.
In contrast, the BS-user direct channel $h^{\left(n\right)}_\mathrm{d}$, which is due to the superimposition of all the non-IRS-refracted paths, generally varies over different blocks in both amplitude and phase (i.e., fast fading), which thus need to be estimated in each block instantaneously.

Based on the above, we can exploit the LoS-dominant channel of the BS-IRS-user link as well as the different behaviors of time-varying channels $\boldsymbol{c}_\mathrm{LoS}^{\left(n\right)}$ and $h^{\left(n\right)}_\mathrm{d}$ to design efficient channel estimation for them. 
Specifically, the user first estimates the effective phases $\{\psi_x,\psi_y\}$ of the LoS-dominant BS-IRS-user cascaded channel, and feeds them back to the IRS for setting its passive beamforming direction to maximize the average channel gain of $h^{\left(n\right)}_\mathrm{r}$ in (\ref{hr}).
Next, the user estimates the resultant IRS-refracted channel $h^{\left(n\right)}_\mathrm{r}$ and the non-IRS-refracted channel $h^{\left(n\right)}_\mathrm{d}$ for subsequent blocks of $n$, and feeds back their phase difference to the IRS for adjusting a common phase shift at all its refracting elements to align these two channels in each block, thus maximizing the received signal power at the user for data transmission.
The above design is expected to achieve both channel estimation and IRS refraction design efficiently. 
Moreover, the proposed IRS refraction design effectively converts the original high-mobility induced fast fading channel (without IRS) to a slow fading counterpart (with IRS), which thus greatly improves the transmission rate and reliability.

\section{Proposed Transmission Protocol and Algorithm Design}
\begin{figure}
\centering
\includegraphics[width=0.7\textwidth]{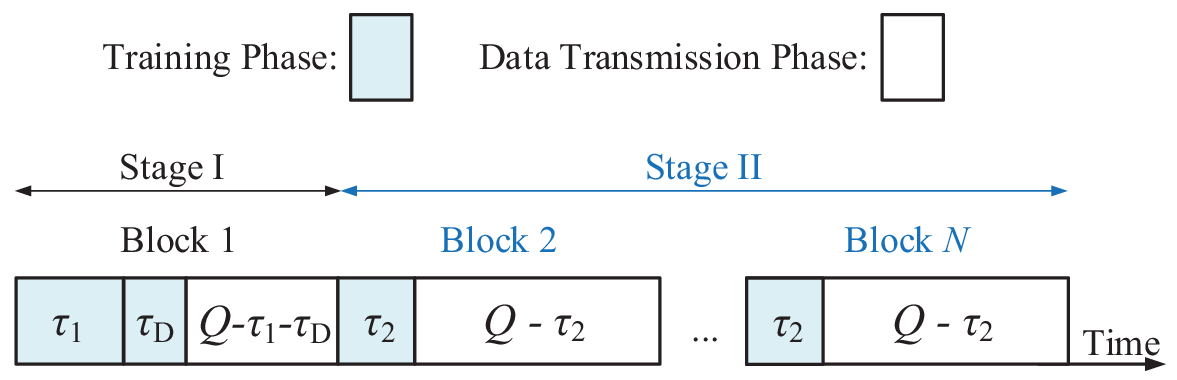}
\caption{Frame structure of the proposed transmission protocol.}
\label{protocol_frame}
\vspace*{-2em}
\end{figure}
In this section, we present the details of our proposed two-stage transmission protocol for the IRS-aided high-mobility communication system, with the frame structure illustrated in Fig.~\ref{protocol_frame}. Specifically, each transmission frame is divided into two stages, referred to as Stage~I and Stage~II, comprising one and $N-1$ block(s), respectively. 
Moreover, each block comprises $Q$ symbols, each with equal duration of $T_s = T_b /Q$.
In the following two subsections, we elaborate the two transmission stages and their pertinent algorithm designs, respectively.
\subsection{Stage~I}
\begin{figure}
\centering
\includegraphics[width=0.4\textwidth]{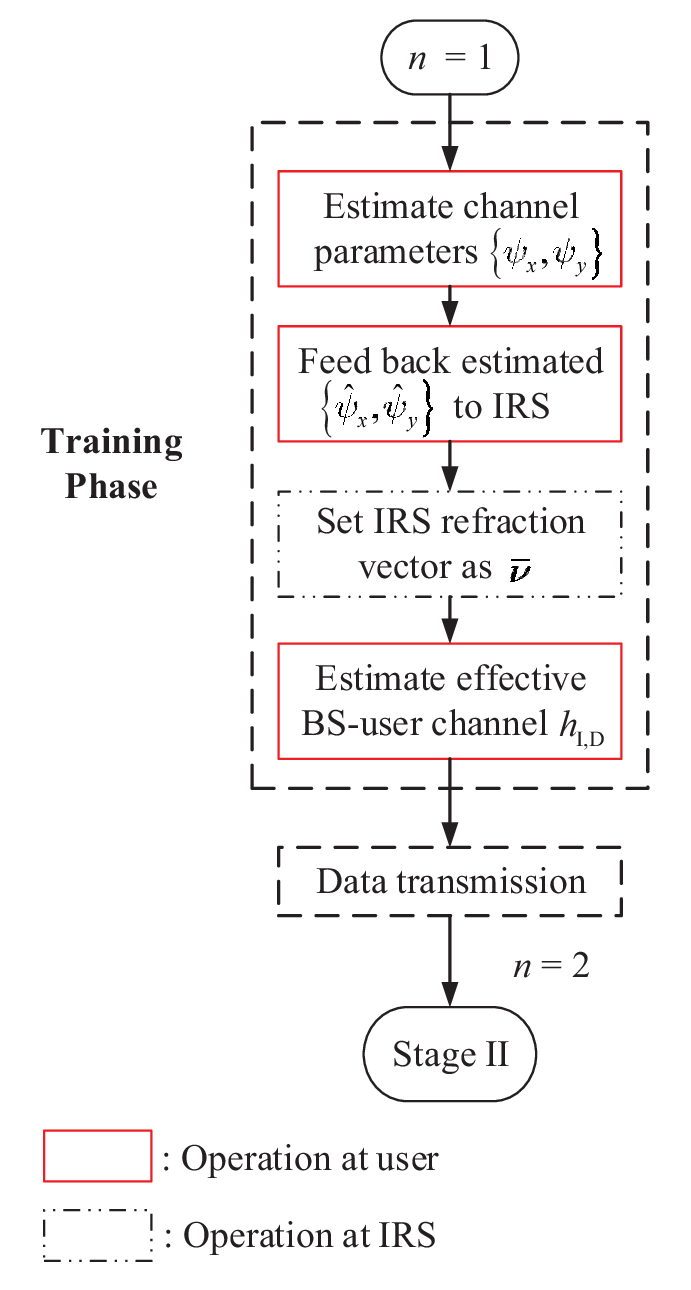}
\caption{Flow chart of Stage~I.}
\label{FC_s1}
\vspace*{-2em}
\end{figure}

As shown by the transmission protocol in Fig. \ref{protocol_frame} and the flow chart of the operations in Stage~I in Fig. \ref{FC_s1}, Stage~I involves only one transmission block which consists of a training phase and a data transmission phase, with $\tau_1+\tau_\mathrm{D}$ and $Q-\tau_1-\tau_\mathrm{D}$ symbols, respectively, where $Q > \tau_1+\tau_\mathrm{D}$. 
During Stage~I, the user estimates the effective phases $\{\psi_x, \psi_y \}$ based on the first $\tau_1$ pilot symbols and feeds them back to the IRS for designing its refraction.
This will help enhance the IRS-refracted channel gain for the subsequent data transmission in Stage~I as well as the channel estimation and data transmission in Stage~II.
After setting the IRS refraction vector according to the effective phases $\{\psi_x, \psi_y \}$, the user then estimates the effective channel for data transmission in Stage~I during the remaining $\tau_\mathrm{D}$ pilot symbol periods.
The details of the two phases in Stage~I are given in the following.
\subsubsection{Training Phase}
Let $\mathcal{T}_\mathrm{I} \triangleq \{1,\ldots,\tau_1 \}$ denote the index set of the first $\tau_1$ pilot symbols sent by the BS in Stage~I. Denote the IRS refraction vector during pilot symbol $i$ as $\boldsymbol{\nu}_i^\mathrm{I}$, where $i \in \mathcal{T}_\mathrm{I}$.
During the training phase of Stage~I, the IRS changes its refraction vector $\boldsymbol{\nu}_i^\mathrm{I}$ over different pilot symbols to facilitate the estimation of the effective phases $\{\psi_x,\psi_y\}$ of the cascaded BS-IRS-user channel. 
Based on the channel model in (\ref{equiva}), the received signal at the user during pilot symbol $i$ can be expressed as
\begin{align}\label{received_sym1}
y_i 
&=  \left( \left(\boldsymbol{\nu}_i^\mathrm{I}\right)^T \boldsymbol{c}^{\left(1\right)}   + h^{\left(1\right)}_\mathrm{d}\right) x_i + z_i \nonumber\\
&=   \left(\boldsymbol{\nu}_i^\mathrm{I}\right)^T \left(\boldsymbol{c}_\mathrm{LoS}^{\left(1\right)} +\boldsymbol{c}_\mathrm{NLoS}^{\left(1\right)}\right)   + h^{\left(1\right)}_\mathrm{d} + z_i \nonumber\\
&=  \beta^{\left(1\right)} \left(\boldsymbol{\nu}_i^\mathrm{I}\right)^T \boldsymbol{u}\left(\psi_x,\psi_y\right)    + h^{\left(1\right)}_\mathrm{d} + \underbrace{\left(\boldsymbol{\nu}_i^\mathrm{I}\right)^T\boldsymbol{c}_\mathrm{NLoS}^{\left(1\right)} + z_i}_{\epsilon_i}
,\quad  i\in \mathcal{T}_\mathrm{I},
\end{align}
where $x_i$ denotes the pilot symbol transmitted and is set as $x_i = 1$, $\forall i$ for ease of exposition,  $z_i \sim\mathcal{N}_{c}\left(0, \sigma^{2}\right)$ is the noise at the user receiver with $\sigma^{2}$ being the normalized noise power (with respect to transmit power), and $\epsilon_i = \left(\boldsymbol{\nu}_i^\mathrm{I}\right)^T\boldsymbol{c}_\mathrm{NLoS}^{\left(1\right)} + z_i$ denotes the equivalent interference-plus-noise term. By letting $\boldsymbol{y} = \left[y_1,\ldots,y_{\tau_1} \right]^T$, the received signal vector over $\tau_1$ pilot symbols can be expressed as 
\begin{align}\label{collect}
\boldsymbol{y}
&=  \beta^{\left(1\right)} \boldsymbol{V} \boldsymbol{u}\left(\psi_x,\psi_y\right)    + h^{\left(1\right)}_\mathrm{d} \boldsymbol{1}_{\tau_1}  +
\boldsymbol{\epsilon},
\end{align}
where $\boldsymbol{V} = \left[\boldsymbol{\nu}_1^\mathrm{I},\ldots,\boldsymbol{\nu}_{\tau_1}^\mathrm{I} \right]^T\in \mathbb{C}^{\tau_1 \times M}$ denotes the IRS refraction matrix and $\boldsymbol{\epsilon} = \left[\epsilon_1,\ldots,\epsilon_{\tau_1}\right]^T\in \mathbb{C}^{\tau_1 \times 1}$ denotes the equivalent interference-plus-noise vector. 
Based on (\ref{collect}), the maximum likelihood (ML) estimation of all relevant unknown channel parameters is given by (with irrelevant terms omitted)
\begin{equation}\label{ls_s1}
\{\hat{\beta}^{\left(1\right)},\hat{\psi}_x,\hat{\psi}_y, \hat{h}_\mathrm{d}^{\left(1\right)} \} = \arg \min _{\beta^{\left(1\right)}, \psi_x,\psi_y, h_\mathrm{d}^{\left(1\right)}} \left\|\boldsymbol{y} -\beta^{\left(1\right)} \boldsymbol{V} \boldsymbol{u}\left(\psi_x,\psi_y\right) - h^{\left(1\right)}_\mathrm{d} \boldsymbol{1}_{\tau_1}\right\|.
\end{equation}
Since there are four unknown parameters in (\ref{ls_s1}), $\tau_1 \ge \operatorname{rank}\left( \boldsymbol{V} \right) \ge 4$ is generally required for the ML estimation in (\ref{ls_s1}). Next, we design the training refraction matrix $\boldsymbol{V}$ in the following two ways:
\begin{enumerate}
    \item \textbf{Random Refraction Design}: The training refraction matrix $\boldsymbol{V}$ is designed as a random matrix with the phase shift of each entry randomly generated from the uniform distribution within $[0,2\pi)$.
    \item \textbf{Discrete Fourier Transform (DFT)-based Refraction Design \cite{ofdmb1}}:
    We draw $l_x$ and $\tau_1 / l_x$ columns (assume $\tau_1 / l_x$ is an integer for convenience) from the $M_x \times M_x$ and $M_y \times M_y$ DFT matrices (denoted by $\boldsymbol{D}_{M_x}$ and $\boldsymbol{D}_{M_y}$, respectively) with semi-equal separation, respectively,  and construct the refraction matrix $\boldsymbol{V}$ as 
    \begin{equation}\label{selected_dft}
    \boldsymbol{\nu}_i^\mathrm{I} = \left[\boldsymbol{D}_{M_x}\right]_{ :,\left\lfloor \frac{\left\lfloor i/l_x \right\rfloor   M_x}{l_x} \right\rfloor  +1}
    \otimes \left[\boldsymbol{D}_{M_y}\right]_{ :,\left\lfloor \frac{M_y l_x\left(\left(i \operatorname{mod} l_x\right)-1\right)}{\tau_1} \right\rfloor  +1}
    ,\quad \forall i\in \mathcal{T}_\mathrm{I}.
\end{equation}
\end{enumerate} 
The performance of the above two training refraction designs will be compared by simulation in Section~IV. 
With fixed $\boldsymbol{V}$, note that given  $\{\beta^{\left(1\right)},\psi_x,\psi_y\}$, the optimal value of $h_\mathrm{d}^{\left(1\right)}$ to minimize the objective function in (\ref{ls_s1}) is given by
\begin{equation}\label{ls_hd}
 \hat{h}_\mathrm{d}^{\left(1\right)} =\frac{\boldsymbol{1}^T_{\tau_1}\left(\boldsymbol{y}- \beta^{\left(1\right)} \boldsymbol{V} \boldsymbol{u}\left(\psi_x,\psi_y\right) \right)}{\tau_1}.
\end{equation}
Substituting (\ref{ls_hd}) into (\ref{ls_s1}), the estimates of $\{\beta^{\left(1\right)}, \psi_x,\psi_y \}$ are given by
\begin{equation}\label{ls_s1_rest2}
\{\hat{\beta}^{\left(1\right)},\hat{\psi}_x,\hat{\psi}_y \} = \arg \min _{\beta^{\left(1\right)}, \psi_x,\psi_y} \left\| \underbrace{\left(\boldsymbol{\mathrm{I}}_{\tau_1} - \frac{\boldsymbol{1}_{\tau_1} \boldsymbol{1}^T_{\tau_1}}{\tau_1} \right)}_{\boldsymbol{B}}  \left(\boldsymbol{y} -\beta^{\left(1\right)} \boldsymbol{V} \boldsymbol{u}\left(\psi_x,\psi_y\right) \right)\right\|.
\end{equation}
For notational convenience, we define $\boldsymbol{\xi}\left(\psi_x,\psi_y\right)= \boldsymbol{B}\boldsymbol{V} \boldsymbol{u}\left(\psi_x,\psi_y\right)$.
For given $\{\psi_x,\psi_y\}$, the optimal value of $\beta^{\left(1 \right)}$ to minimize the objective function in (\ref{ls_s1_rest2}) is given by
\begin{align}\label{ls_s1_rest3}
\hat{\beta}^{\left( 1 \right)} 
=\frac{\boldsymbol{\xi}^H\left(\psi_x,\psi_y\right)  \boldsymbol{B}\boldsymbol{y}}{\left\| \boldsymbol{\xi}\left(\psi_x,\psi_y\right) \right\|^2}
\stackrel{(b)}{=}\frac{\boldsymbol{\xi}^H\left(\psi_x,\psi_y\right)  \boldsymbol{y}}{\left\| \boldsymbol{\xi}\left(\psi_x,\psi_y\right) \right\|^2}
,
\end{align}
where $(b)$ holds due to the fact that $\boldsymbol{B}^H\boldsymbol{B} = \boldsymbol{B}$ and thus $\boldsymbol{\xi}^H\left(\psi_x,\psi_y\right)  \boldsymbol{B} = \boldsymbol{\xi}^H\left(\psi_x,\psi_y\right) $.
Substituting (\ref{ls_s1_rest3}) into (\ref{ls_s1_rest2}), the ML estimates of $\{\psi_x,\psi_y\}$ are given by
\begin{equation}\label{ls_s1_psi}
\{\hat{\psi}_x,\hat{\psi}_y\} = \arg \max _{\psi_x,\psi_y} 
\frac{
\left| \boldsymbol{\xi}^H \left(\psi_x,\psi_y\right) \boldsymbol{y}
 \right|^2
}{\left\| \boldsymbol{\xi}\left(\psi_x,\psi_y\right) \right\|^2}
.
\end{equation}
It can be verified that the problem in (\ref{ls_s1_psi}) is a non-convex optimization problem as its objective function is non-concave with respect to $\psi_x$ and $\psi_y$, which is thus difficult to be solved optimally. 
As such, we propose an efficient two-step algorithm to solve (\ref{ls_s1_psi}) sub-optimally as follows.
\begin{itemize}[leftmargin=*]
    \item \textbf{Step~1}: Let $\{A_x,A_y\}$ denote the number of quantization levels for $\{\psi_x,\psi_y\}$, respectively. Consider the uniform quantization as follows
    \begin{equation}
    \begin{aligned}\label{gp}
    \mathcal{G} \triangleq\Big\{\{\psi_x^\jmath,\psi_y^\kappa\} \Bigr| \psi_x^\jmath &=-1+\frac{2\jmath}{A_x},\jmath = 1,\ldots,A_x, \\
    \psi_y^\kappa &=-1+\frac{2\kappa}{A_y},\kappa = 1,\ldots,A_y\Big\}.
    \end{aligned}
    \end{equation}
    Accordingly, an exhaustive 2D grid-based search over $\mathcal{G}$ can be performed to find an initial solution to (\ref{ls_s1_psi}), which is given by
    \begin{equation}\label{gs_solu}
    \{\psi_x^\mathrm{GS},\psi_y^\mathrm{GS}\} = \arg \max _{\{\psi_x,\psi_y\} \in \mathcal{G}} \frac{\left| \boldsymbol{\xi}^H \left(\psi_x,\psi_y\right)\boldsymbol{y}\right|^2}{\left\| \boldsymbol{\xi}\left(\psi_x,\psi_y\right) \right\|^2}
.
    \end{equation}
    \item \textbf{Step~2}:
    With $\{\psi_x^\mathrm{GS},\psi_y^\mathrm{GS}\}$ obtained in Step~1, we then refine the estimates by further applying a simple gradient-based search algorithm \cite{cvx_by} to obtain the final estimates $\{\hat{\psi}_x,\hat{\psi}_y\}$, which is summarized in Algorithm~1. 
    Moreover, Algorithm 1 is terminated either when the maximum number of iterations $I_\mathrm{max}$ is reached or the difference between any two consecutive iterations is smaller than a small positive threshold, denoted by $\varrho$.
    For the derivation of the search direction $\boldsymbol{\Delta }\left(\psi_x,\psi_y\right)$ in Algorithm~1, please refer to Appendix A.
    \vspace*{-1em}
    \begin{algorithm}
    \SetAlgoLined
    \SetEndCharOfAlgoLine{}
    \textbf{Input}: $\psi_x^\mathrm{GS},\psi_y^\mathrm{GS}$;
    \textbf{Output}: $\hat{\psi}_x,\hat{\psi}_y$ ;
 
    set $\{\hat{\psi}_x,\hat{\psi}_y\} =\{\psi_x^\mathrm{GS},\psi_y^\mathrm{GS}\}$;
 
    \Repeat{ 
    {\upshape the obtained}  $\hat{\psi}_x,\hat{\psi}_y$ {\upshape  reach convergence.}
    } 
    {Compute the search direction: $\boldsymbol{\Delta }\left(\hat{\psi}_x,\hat{\psi}_y\right)=\nabla \frac{\left| \boldsymbol{\xi}^H \left(\hat{\psi}_x,\hat{\psi}_y\right)\boldsymbol{y}\right|^2}{\left\| \boldsymbol{\xi}\left(\hat{\psi}_x,\hat{\psi}_y\right) \right\|^2}$;

    Choose step size $t$ via the backtracking line search \cite{cvx_by} ;
 
    Update: $\left[\hat{\psi}_x,\hat{\psi}_y\right]^T = \left[\hat{\psi}_x,\hat{\psi}_y\right]^T+ t \boldsymbol{\Delta }\left(\hat{\psi}_x,\hat{\psi}_y\right)$ ;}
    \caption{Gradient-based Search}
    \end{algorithm}
    \vspace*{-2em}
\end{itemize}

The complexity of solving (\ref{gs_solu}) in Step~1 is $\mathcal{O}\left(\tau_1 A_x A_y \right)$ and that of Algorithm~1 in Step~2 is $\mathcal{O}(\tau_1 I_\mathrm{it} )$, where $I_\mathrm{it}$ denotes the number of iterations required for convergence. Thus, the overall complexity for the proposed two-step algorithm is given by $\mathcal{O}\left(\tau_1 \left(A_x A_y+I_\mathrm{it}\right) \right)$.

After the above estimation, the user feeds back $\{\hat{\psi}_x,\hat{\psi}_y\}$ to the IRS controller via a separate wireless control link. For simplicity, we assume that the control link is reliable with negligible delay\footnote{The proposed protocol can be modified to accommodate small but non-negligible feedback delay, for which the performance will be investigated by simulation in Section~IV.}. 
Based on the feedback information $\{\hat{\psi}_x,\hat{\psi}_y\}$ from the user, the IRS sets its refraction vector as
\begin{align}\label{init_refrac}
\bar{\boldsymbol{\nu}} = \boldsymbol{u}^*\left(\hat{\psi}_x,\hat{\psi}_y\right), 
\end{align} 
to maximize the LoS path gain of the IRS-refracted channel.
With the IRS refraction vector set as (\ref{init_refrac}), the BS then sends additional $\tau_\mathrm{D} \ge 1$ pilot symbols to the user for estimating the effective channel for data transmission in Stage~I. Let $\mathcal{T}_\mathrm{D} \triangleq \{\tau_1 +1,\ldots, \tau_1+\tau_\mathrm{D} \}$ denote the index set for the additional $\tau_\mathrm{D}$ pilot symbols. With $x_i = 1$, $\forall i \in \mathcal{T}_\mathrm{D}$ being the pilot symbol transmitted by the BS, the received signal can be expressed as 
\begin{align}\label{s1_dest_sym}
y_i = \underbrace{ \left(\bar{\boldsymbol{\nu}}\right)^T \boldsymbol{c}^{\left(1\right)}   + h^{\left(1\right)}_\mathrm{d}}_{h_\mathrm{I,D}}  + z_i, \quad \forall i \in \mathcal{T}_\mathrm{D},
\end{align} 
where $h_\mathrm{I,D}$ denotes the effective channel for data transmission and $z_i \sim\mathcal{N}_{c}\left(0, \sigma^{2}\right)$ is the noise at the user receiver. Based on (\ref{s1_dest_sym}), the effective channel $h_\mathrm{I,D}$ is estimated as 
\begin{align}\label{s1_dest}
\hat{h}_\mathrm{I,D} = \frac{1}{\tau_\mathrm{D}} \sum_{i\in \mathcal{T}_\mathrm{D}} y_i.
\end{align}

\subsubsection{Data Transmission Phase}
Denote the IRS refraction vector for data transmission in Stage~I as $\boldsymbol{\nu}_\mathrm{I,D}$.
With the IRS refraction vector set as $\boldsymbol{\nu}_\mathrm{I,D}=\bar{\boldsymbol{\nu}}$ in (\ref{init_refrac}) and based on the estimated CSI of $\hat{h}_\mathrm{I,D}$ in (\ref{s1_dest}), the user decodes the data in the remaining $Q-\tau_1-\tau_\mathrm{D}$ symbols of Stage~I. Thus, the achievable rate (with the training overhead taken into account) in bits
per second per Hertz (bps/Hz) of Stage~I is given by
\begin{align}\label{rate1}
R_\mathrm{I} = \frac{Q -\tau_1-\tau_\mathrm{D}}{Q} \log _{2}\left(1+\frac{ W_\mathrm{I} }{\Gamma \sigma^{2}}\right), 
\end{align}
where $W_\mathrm{I}  = \left| \boldsymbol{\nu}_\mathrm{I,D}^T
\boldsymbol{c}^{\left(1\right)} + h^{\left(1\right)}_\mathrm{d} \right|^2$ denotes the effective channel power gain of Stage~I for data transmission
and $\Gamma \ge 1$ denotes the achievable rate gap from the channel capacity \cite{gap} due to the practical modulation and coding scheme used.

\subsection{Stage~II}
\begin{figure}
\centering
\includegraphics[width=0.45\textwidth]{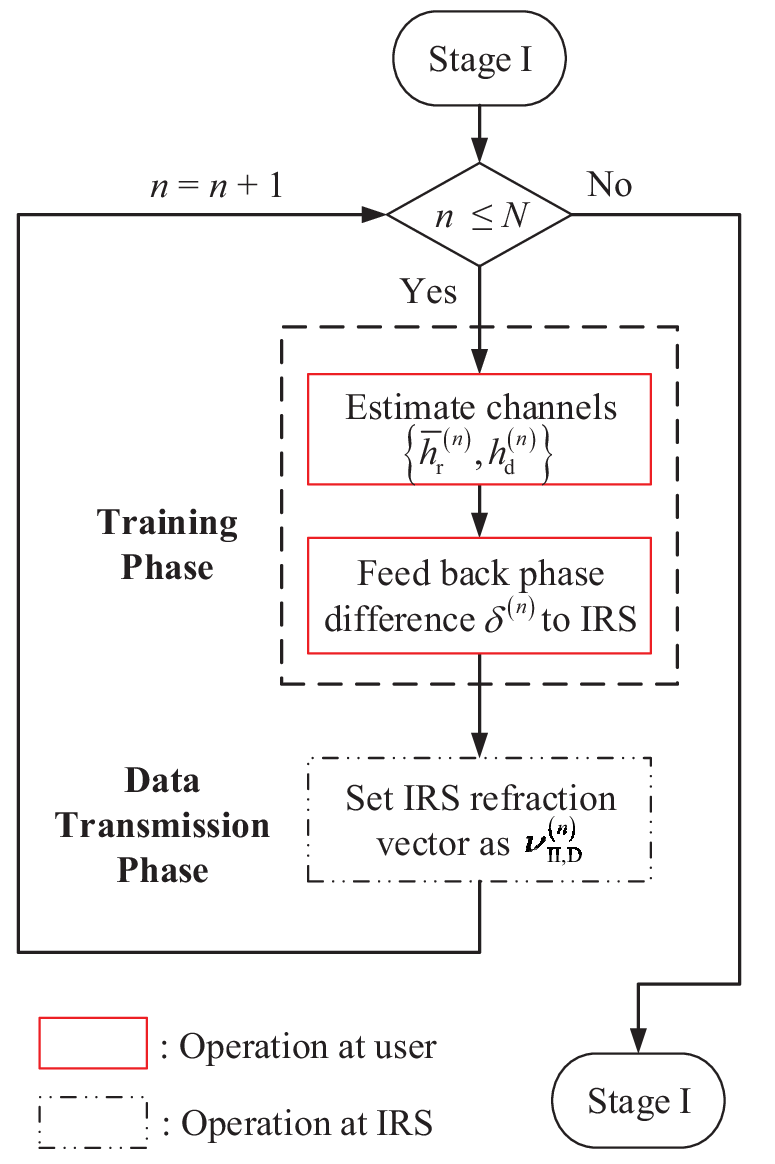}
\caption{Flow chart of Stage~II.}
\label{FC_s2}
\vspace*{-2em}
\end{figure}

As shown in the transmission protocol in Fig. \ref{protocol_frame} and the flow chart of the operations in Stage~II in Fig. \ref{FC_s2}, Stage~II consists of $N-1$ blocks, where each block comprises a training phase and a data transmission phase, with $\tau_2$ and $Q-\tau_2$ symbols, respectively, where $Q > \tau_2$. 
With the IRS refraction vector set as  $\bar{\boldsymbol{\nu}}$ in (\ref{init_refrac}) that maximizes the LoS path gain of the IRS-refracted channel, we aim to subsequently maximize the overall BS-user channel gain in each block of Stage~II, by coherently combining the IRS-refracted channel with the non-IRS-refracted channel.
In the following, we elaborate the operations and algorithm designs for each block in Stage~II for channel training and data transmission, respectively.
\subsubsection{Training Phase}
Let $\mathcal{N}_\mathrm{II} \triangleq \{2,\ldots,N \}$ denote the index set of the $N-1$  blocks in Stage~II.
To maximize the end-to-end BS-user channel gain for data transmission with the coherent combination of the IRS-refracted and non-IRS-refracted channels, we need to estimate these two (scalar) channels during the training phase of each block $n \in \mathcal{N}_\mathrm{II}$.
Let $\mathcal{T}_\mathrm{II} \triangleq \{1,\ldots,\tau_2 \}$ denote the index set for $\tau_2$ pilot symbols of each block in Stage~II.
The IRS tunes its refractions over $\tau_2$ pilot symbols for each block in Stage~II according to
\begin{align}\label{refrac_tra}
\boldsymbol{\nu}_i^\mathrm{II} =  \mu_i \bar{\boldsymbol{\nu}}
,\quad  i\in \mathcal{T}_\mathrm{II},
\end{align}
where $ \mu_i \in \mathbb{C} $ with $ \left|\mu_i\right| =1 $ denotes the common phase shift that applies to all the refracting elements.
As such, with $x^{\left(n\right)}_i = 1$ being the pilot symbol transmitted by the BS, the received signal at the user during pilot symbol $i$ of block $n$ can be expressed as
\begin{align}\label{received_sym2}
y^{\left(n\right)}_i 
&=  \left(\boldsymbol{\nu}_i^\mathrm{II}\right)^T \boldsymbol{c}^{\left(n\right)}   + h^{\left(n\right)}_\mathrm{d} + z^{\left(n\right)}_i
\nonumber\\
&= \mu_i \underbrace{\bar{\boldsymbol{\nu}}^T  \boldsymbol{c}^{\left(n\right)}}_{\bar{h}^{\left(n\right)}_\mathrm{r}}   + h^{\left(n\right)}_\mathrm{d} + z^{\left(n\right)}_i
,\quad n\in \mathcal{N}_\mathrm{II}, i\in \mathcal{T}_\mathrm{II},
\end{align}
where $\bar{h}^{\left(n\right)}_\mathrm{r} = \bar{\boldsymbol{\nu}}^T  \boldsymbol{c}^{\left(n\right)}$ denotes the initial IRS-refracted channel in block $n$ and $z^{\left(n\right)}_i \sim\mathcal{N}_{c}\left(0, \sigma^{2}\right)$ is the received noise.
By stacking $\tau_2$ received pilot symbols $\{y^{\left(n\right)}_i\}_{i=1}^{\tau_2}$ of each block into $\boldsymbol{y}^{\left(n\right)} = \left[y_1^{\left(n\right)},\ldots,y_{\tau_2}^{\left(n\right)} \right]^T$, the received signal vector can be expressed as 
\begin{equation}\label{received_ce}
\boldsymbol{y}^{\left(n\right)} = \boldsymbol{\Theta} \boldsymbol{h}^{\left(n\right)} + \boldsymbol{z}^{\left(n\right)}, \quad n\in \mathcal{N}_\mathrm{II},
\end{equation}
where $\boldsymbol{\Theta} = \left[ \bar{\boldsymbol{\mu}}_1, \ldots,  \bar{\boldsymbol{\mu}}_{\tau_2}  \right]^T \in \mathbb{C}^{\tau_2 \times 2}$ denotes the common training refraction matrix in Stage~II with $\bar{\boldsymbol{\mu}}_i = \left[1, \mu_i\right]^T $, $\boldsymbol{h}^{\left(n\right)} = \left[ h^{\left(n\right)}_{\scaleto{\mathrm{d}}{5.5pt}}, \bar{h}^{\left(n\right)}_{\scaleto{\mathrm{r}}{4.8pt}}\right]^T$ denotes the channel vector including both the IRS-refracted and non-IRS-refracted channels, 
and $\boldsymbol{z}^{\left(n\right)} = \left[z^{\left(n\right)}_{1}, \ldots, z^{\left(n\right)}_{\tau_2}\right]^{T}$ denotes the  received noise vector. 
By properly constructing the training matrix $\boldsymbol{\Theta}$ such that $\mathrm{rank} \left(\boldsymbol{\Theta}\right) = 2$, the least squares (LS) estimate of $\boldsymbol{h}^{\left(n\right)}$ at the user based on (\ref{received_ce}) is given by
\begin{equation}\label{est_bl}
\hat{\boldsymbol{h}}^{\left(n\right)} 
=\boldsymbol{\Theta}^\dagger
\boldsymbol{y}^{\left(n\right)} = \boldsymbol{h}^{\left(n\right)}  + \boldsymbol{\Theta}^\dagger \boldsymbol{z}^{\left(n\right)}, \quad n\in \mathcal{N}_\mathrm{II}.
\end{equation}
Note that $\tau_2 \ge 2$ is required to ensure $\mathrm{rank} \left(\boldsymbol{\Theta}\right) = 2$ and thus the existence of $ {\boldsymbol \Theta}^{\dagger}$. 
For example, the IRS can set the training matrix $\boldsymbol{\Theta}$ as the sub-matrix of the $\tau_2\times  \tau_2$ DFT matrix with its first two columns.
Then, the phase difference of the estimated IRS-refracted and non-IRS-refracted channels, denoted by 
\begin{equation}\label{phase_difference}
\delta^{\left(n\right)} = -\angle \hat{\bar{h}}^{\left(n\right)}_\mathrm{r} + \angle \hat{h}^{\left(n\right)}_\mathrm{d},\quad n\in \mathcal{N}_\mathrm{II},
\end{equation}
is fed back to the IRS for refining its refraction for data transmission (to be specified next).

\subsubsection{Data Transmission Phase}
Denote the IRS refraction vector for data transmission in block $n$ as $\boldsymbol{\nu}^{\left(n\right)}_\mathrm{II,D} = \mu_\mathrm{II,D}^{\left(n\right)} \boldsymbol{\bar{\nu}}$, where $\mu_\mathrm{II,D}^{\left(n\right)}$ with $ \left|\mu_\mathrm{II,D}^{\left(n\right)}\right| =1 $ denotes the common phase shift that applies to all the refracting elements.
Based on the estimated $\hat{\boldsymbol{h}}^{\left(n\right)}$ in (\ref{est_bl}), the estimated effective channel for data transmission in block $n$ is given by
\begin{equation}\label{w2}
\hat{h}^{\left(n\right)}_\mathrm{II,D}  =\mu_\mathrm{II,D}^{\left(n\right)} \hat{\bar{h}}^{\left(n\right)}_\mathrm{r}  + \hat{h}^{\left(n\right)}_\mathrm{d} , \quad n\in \mathcal{N}_\mathrm{II}.
\end{equation}
Based on (\ref{w2}), we have
\begin{equation}\label{cgg}
\left|\hat{h}^{\left(n\right)}_\mathrm{II,D}  \right|
= \left|\mu_\mathrm{II,D}^{\left(n\right)} \hat{\bar{h}}^{\left(n\right)}_\mathrm{r}  + \hat{h}^{\left(n\right)}_\mathrm{d}\right| 
\stackrel{(c)}{\leq} \left| \hat{\bar{h}}^{\left(n\right)}_\mathrm{r}\right| + \left|\hat{h}^{\left(n\right)}_\mathrm{d}\right|
, \quad n\in \mathcal{N}_\mathrm{II},
\end{equation}
where $(c)$ is due to the triangle inequality and the equality holds if and only if $\angle \left( \mu_\mathrm{II,D}^{\left(n\right)} \hat{\bar{h}}^{\left(n\right)}_\mathrm{r} \right) = \angle \hat{h}^{\left(n\right)}_\mathrm{d}$. Thus the common phase shift $\mu_\mathrm{II,D}^{\left(n\right)}$ at the IRS is designed as
\begin{equation}\label{v_design}
\mu_\mathrm{II,D}^{\left(n\right)}
= e^{j \delta^{\left(n\right)} } , \quad n\in \mathcal{N}_\mathrm{II},
\end{equation}
so as to coherently combine the IRS-refracted channel with the non-IRS-refracted channel.
With the training overhead of Stage~II taken into account, the achievable rate of Stage~II is given by 
\begin{align}\label{rate2}
R_\mathrm{II} = \frac{Q -\tau_2}{\left(N-1\right) Q} \sum_{n= 2}^{N} \log _{2}\left(1+\frac{ W_\mathrm{II}^{\left(n\right)}}{\Gamma \sigma^{2} }\right),
\end{align}
where $W_\mathrm{II}^{\left(n\right)} = \left|\left( \boldsymbol{\nu}^{\left(n\right)}_\mathrm{II,D} \right)^{T} \boldsymbol{c}^{\left(n\right)}+h^{\left(n\right)}_\mathrm{d} \right|^2$ denotes the effective channel power gain of block $n$ in Stage~II for data transmission. 
Note that owing to the IRS passive beamforming gain and coherent signal combining at the user, $W_\mathrm{II}^{\left(n\right)}$ in Stage~II is expected to be significantly larger than $W_\mathrm{I}$ in Stage~I, as will be shown by simulation in Section~IV.  
By integrating the achievable rates given in (\ref{rate1}) and (\ref{rate2}) for Stages~I~and~II, respectively, the overall achievable rate of one transmission frame is given by
\begin{align}\label{rateover}
R =  \frac{1}{N}  R_\mathrm{I} + \frac{N -1}{N}  R_\mathrm{II}.
\end{align}

\section{Simulation Results}
In this section, we evaluate the performance of our proposed on-vehicle IRS aided high-mobility communication system and two-stage transmission protocol as well as various algorithms by simulation.
We set the carrier frequency as $f_c = 5.9$~GHz, and the signal bandwidth as 500~KHz. 
The vehicle speed is set as $v=50$~m/s (if not specified otherwise), which results in a Doppler frequency with the maximum value of $f_{max} = v f_c/c  \approx 1 $~KHz, where $c = 3 \times 10^8$~m/s denotes the speed of light.
Each frame consists of $N = 40$ blocks and the duration of each block is set as $T_b = 1/\left(5 f_{max}\right) \approx 0.2$~millisecond (ms), during which the BS-user direct/BS-IRS channel is assumed to remain approximately constant. 
We consider different values of $M$ in our simulations, by fixing $M_y = 10$ and changing $M_x$ linearly with $M$.
We set the half-wavelength spacing for adjacent IRS refracting elements.
We set the BS-IRS, BS-user, and IRS-user distances as 100~m, 100~m, and 2~m, respectively, unless otherwise stated.
The path loss exponents of the BS-user, BS-IRS, IRS-user links are set as 3, 2.2, and 2.2, respectively, to cater to their different link distances as well as LoS availability, and the channel power gain at the reference distance of 1~m is set as $\varepsilon_0 = -30$~dB for each link. 
The AoAs of $\{\theta_\mathrm{BI},\vartheta_\mathrm{BI}\}$ and AoDs $\{\theta_\mathrm{IU},\vartheta_\mathrm{IU}\}$ are randomly generated from the uniform distribution within their respective ranges defined in Section~II, unless otherwise stated.
We assume that the BS-user direct channel $h^{\left(n\right)}_\mathrm{d}$ for $n\in \mathcal{N}$, and the NLoS component of the BS-IRS channel follow the Rayleigh fading with their time correlation following the Jake's spectrum \cite{gold}.
Let $P_t$ denote the transmit power at the BS and the noise power at the user is set as $\sigma_u^2 = -110$~dBm. Accordingly, the normalized noise power at the user is given by $\sigma^2 = \sigma_u^2/P_t$.
For the quantization levels of the 2D grid-based search for $\{\psi_x, \psi_y\}$ in Stage~I, we set $A_x = A_y = 20$.
For the training phase in Stage~I, we set $\tau_\mathrm{D} = 1$.
For each block in Stage~II, we set $\tau_2 = 2$ and the gap to channel capacity is set as $\Gamma = 9$~dB. 

\subsection{Channel Parameter Estimation in Stage~I}
\subsubsection{Performance of IRS Training Refraction Design based on Estimated Channel}
\begin{figure} 
     \begin{subfigure}{0.49\textwidth}
         \centering
         \includegraphics[width=\textwidth]{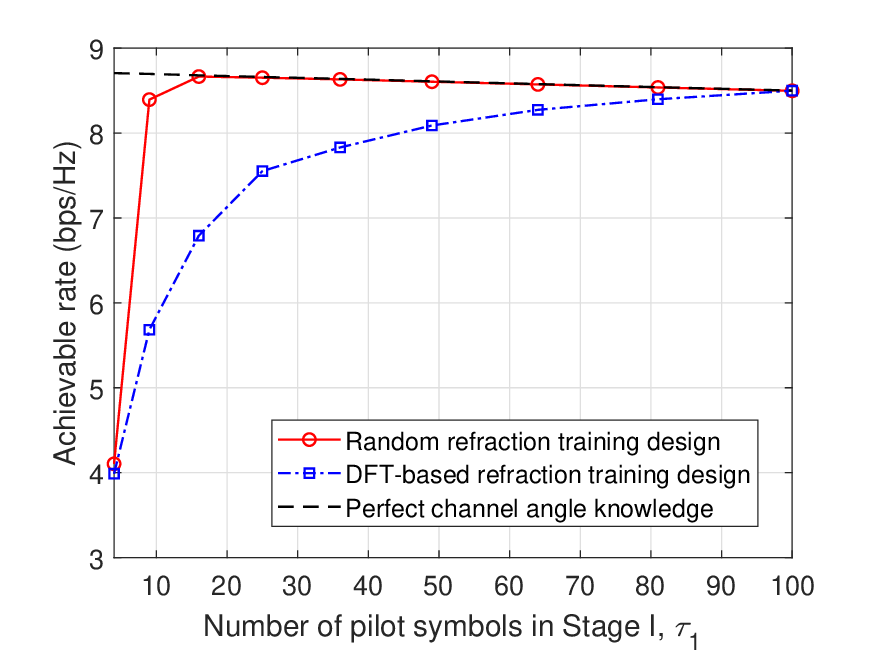}
         \caption{Achievable rate $R$ versus $\tau_1$.}
     \end{subfigure}
     \hfill
     \begin{subfigure}{0.49\textwidth}
         \centering
         \includegraphics[width=\textwidth]{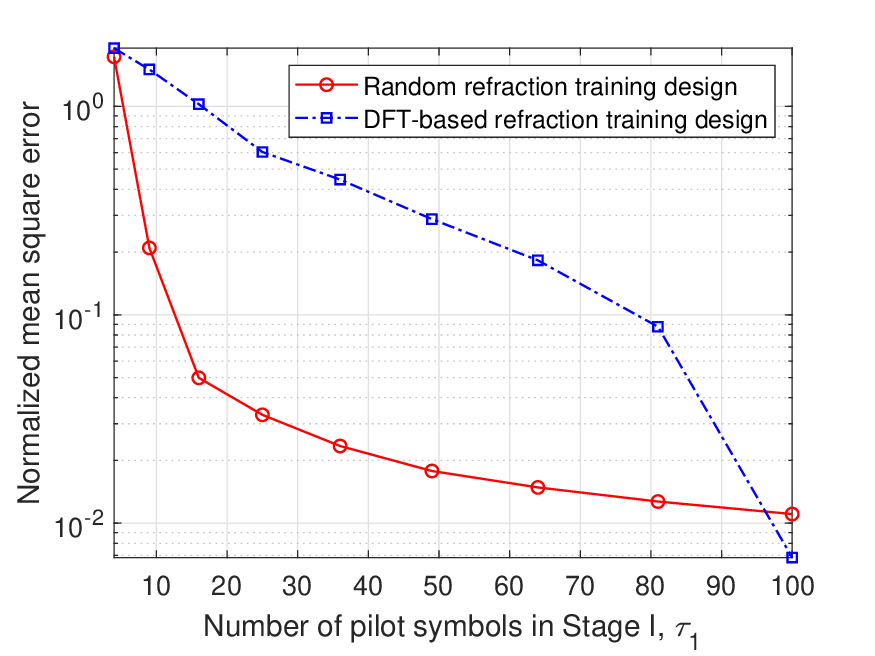}
         \caption{NMSE versus $\tau_1$.}
     \end{subfigure}
     \caption{Performance comparison of different training refraction matrix designs in Stage~I.}
     \label{rand_vs_dft}
     \vspace*{-2em}
\end{figure}
First, we evaluate the performance of our proposed scheme with the two training refraction designs (i.e., random refraction training design and DFT-based refraction training design in (\ref{selected_dft})) for estimating $\{\psi_x,\psi_y\}$ in Stage~I.
We also consider the case with their perfect knowledge as the performance upper bound, where we have $\{\hat{\psi}_x,\hat{\psi}_y\}=\{\psi_x,\psi_y\}$ in (\ref{ls_s1_psi}). 
Define the normalized mean square error (NMSE) of the estimated LoS component of the cascaded BS-IRS-user channel as $\mathbb{E}\left\{\frac{\left\| 
\boldsymbol{u}\left(\psi_x,\psi_y\right)
- \boldsymbol{u}\left(\hat{\psi}_x,\hat{\psi}_y\right)\right\|^2}{\left\| \boldsymbol{u}\left(\psi_x,\psi_y\right)\right\|^2}\right\}$. 
We plot in Fig. \ref{rand_vs_dft}(a) the achievable rate $R$ and in Fig. \ref{rand_vs_dft}(b) the NMSE performance, both versus $\tau_1$, with $P_t = 26$~dBm, $M = 100$, and $K = 10$~dB.
It is observed that the proposed scheme with the random refraction training design achieves a higher rate than that with the DFT-based refraction training design in Fig. \ref{rand_vs_dft}(a). This is in accordance with the NMSE performance in Fig. \ref{rand_vs_dft}(b) and can be explained as follows. The DFT-based refraction training design applies directional beamforming at the IRS over different training symbols, which can result in very low received signal power when $\{\psi_x,\psi_y\}$ is out of the beam coverage and thus degrade the estimation accuracy. In contrast, the random refraction training design prevents the received signal power from significantly varying over different training symbols for any given $\{\psi_x,\psi_y\}$, thus achieving more accurate estimation for a given number of pilot symbols. 
Hence, in the following simulations, the random refraction based training design is adopted for the training phase in Stage~I.
\subsubsection{Performance of Algorithm 1 for the ML Estimation}
\begin{figure}
\centering
\includegraphics[width=0.60\textwidth]{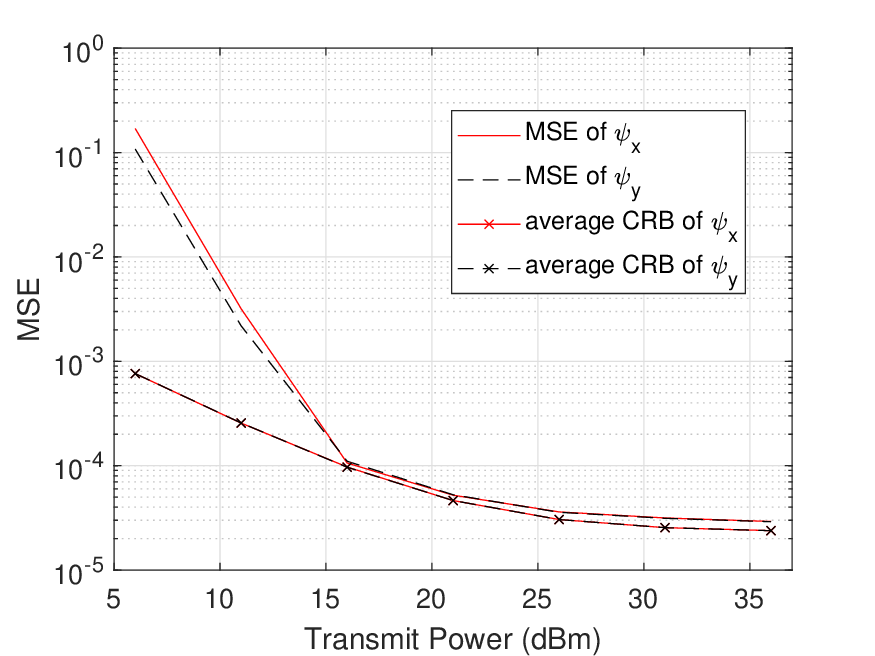}
\caption{MSE versus transmit power $P_t$.}
\label{crb}
\vspace*{-2em}
\end{figure}
We define the mean square error (MSE) of the effective phases $\{\psi_x,\psi_y\}$ as $\mathbb{E}\left\{ \left(\psi_x -\hat{\psi}_x  \right)^2  \right\}$ and $\mathbb{E}\left\{ \left(\psi_y -\hat{\psi}_y  \right)^2  \right\}$, respectively. To evaluate the performance of the ML estimation with random refraction training design, we consider the Cram\'{e}r-Rao bound (CRB) for performance comparison. For the derivation of the CRB, please refer to Appendix B for details. Note that the CRBs of the relevant parameters are in general functions of both the effective phases $\{\psi_x,\psi_y\}$ and the training refraction pattern $\boldsymbol{V}$; thus, the average CRB is taken as the performance bound.
We plot in Fig.~\ref{crb} the MSE versus the transmit power $P_t$, with $\psi_x = 0.5$, $\psi_y = 0.8$, $K = 10$~dB, $M = 100$, and $\tau_1 = 30$. It is observed that the MSE is tightly lower-bounded by the average CRB when $P_t > 15$~dBm, which validates the effectiveness of the proposed two-step estimator in estimating the effective phases.

\subsubsection{Effect of Rician Factor $K$ of the BS-IRS Channel}
\begin{figure}
\centering
\includegraphics[width=0.60\textwidth]{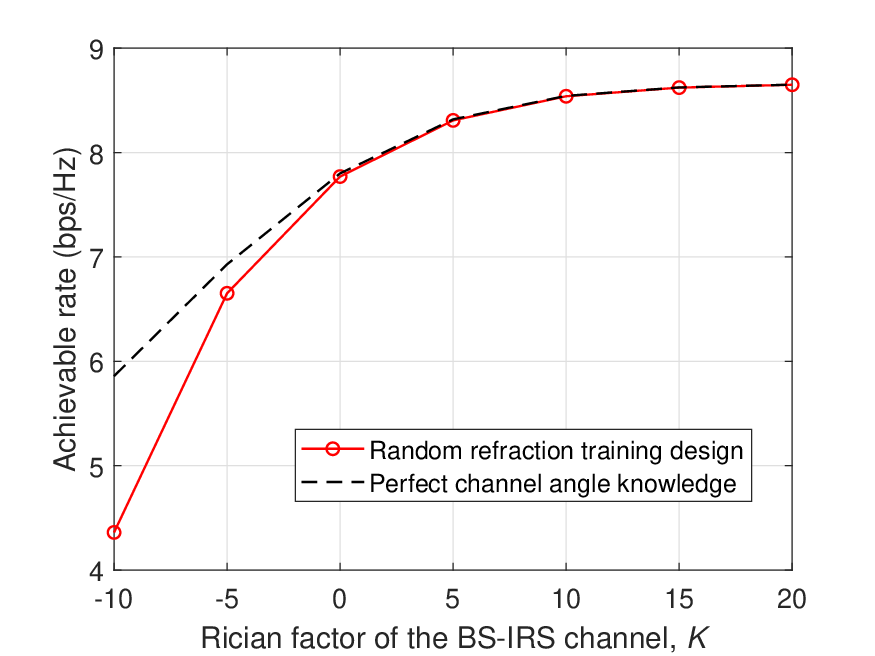}
\caption{Achievable rate $R$ versus $K$.}
\label{rate_rician}
\vspace*{-2em}
\end{figure}
Next, we evaluate the effect of the Rician factor $K$ of the BS-IRS channel.
In Fig.~\ref{rate_rician}, we show the achievable rate $R$ versus the Rician factor $K$ of the BS-IRS channel, with $P_t = 26$~dBm, $M = 100$, and $\tau_1 = 80$.
It is observed that the achievable rates of the two cases (i.e., the proposed scheme with the random refraction design and the case assuming perfect channel angle/phase knowledge) both increase with $K$. This is because the IRS refraction design in Stage~I is aimed to maximize the LoS path gain of the BS-IRS-user cascaded channel and the average power of this LoS path increases with $K$.
Moreover, it is observed that the proposed scheme achieves almost the same rate performance as that with perfect channel phase information when $K \ge  0$~dB, which implies that our proposed scheme is very practically robust against the NLoS component in the BS-IRS-user cascaded channel.

\subsection{Refraction Design for Data Transmission in Stage~II}
Next, we evaluate the performance of the proposed IRS refraction design for data transmission in Stage~II with the following two benchmark designs:

\begin{enumerate}
    \item \textbf{Modified Scheme with Feedback Delay (FD)}: In this case, we take the feedback delay from the user to IRS controller into account for the IRS refraction design. We assume that the IRS can only adjust its refraction at each block based on the user feedback in the previous block due to the feedback delay. To meet this new constraint, our proposed transmission protocol is modified as follows. First, for Stage~I, instead of setting $\boldsymbol{\nu}_\mathrm{I,D} = \bar{\boldsymbol{\nu}}$, the IRS simply sets the refraction vector for data transmission as  $\boldsymbol{\nu}_\mathrm{I,D} = \boldsymbol{\nu}^\mathrm{I}_{\tau_1}$, i.e., the IRS training refraction vector used for the last pilot symbol in (\ref{received_sym1}). As a result, the additional $\tau_\mathrm{D}$ pilot symbols are no more needed (i.e., we set $\tau_\mathrm{D} = 0$), since the effective channel $h_\mathrm{I,D} = \left(\boldsymbol{\nu}^\mathrm{I}_{\tau_1}\right)^T \boldsymbol{c}^{\left(1\right)}   + h^{\left(1\right)}_\mathrm{d}$ for decoding the subsequent data in Stage~I has been estimated at the user as $\hat{h}_\mathrm{I,D}= y_{\tau_1}$.
    Second, for Stage~II, we set the IRS common phase shift for data transmission at block $n$, $n\in \mathcal{N}_\mathrm{II}$, based on the estimated phase difference $\delta^{\left(n-1\right)}$ in (\ref{phase_difference}) from the previous block, which is given by
    \begin{equation}\label{v_design_fd}
    \mu_\mathrm{II,D}^{\left(n\right)}= 
    \begin{cases}
          1, & n =2,\\
          e^{j \delta^{\left(n-1\right)} }, & n\in \{3,\ldots, N\}.
        \end{cases}
    \end{equation}
    \item \textbf{Proposed Scheme without (w/o) Channel Phase Alignment (CPA)}: In this case, to reduce the implementation (for IRS common refraction phase shift and user phase difference feedback in Stage~II) complexity, we only perform the passive beamforming for the IRS-refracted link using $\bar{\boldsymbol{\nu}}$ obtained in Stage~I but without coherently combining it with the non-IRS-refracted link in Stage~II. As such, we set the IRS common phase shift for data transmission in Stage~II as $\mu_\mathrm{II,D}^{\left(n\right)} = 1, \forall n\in \mathcal{N}_\mathrm{II}$. 
    Since the IRS-refracted and non-IRS-refracted channels are no longer needed to be estimated separately as that in (\ref{est_bl}), we set $\tau_2 = 1$ for estimating the effective channel for data transmission as $\hat{h}^{\left(n\right)}_\mathrm{II,D} = y^{\left(n\right)}_1, \forall n\in \mathcal{N}_\mathrm{II}$.
\end{enumerate} 
\begin{figure}
\centering
\includegraphics[width=0.60\textwidth]{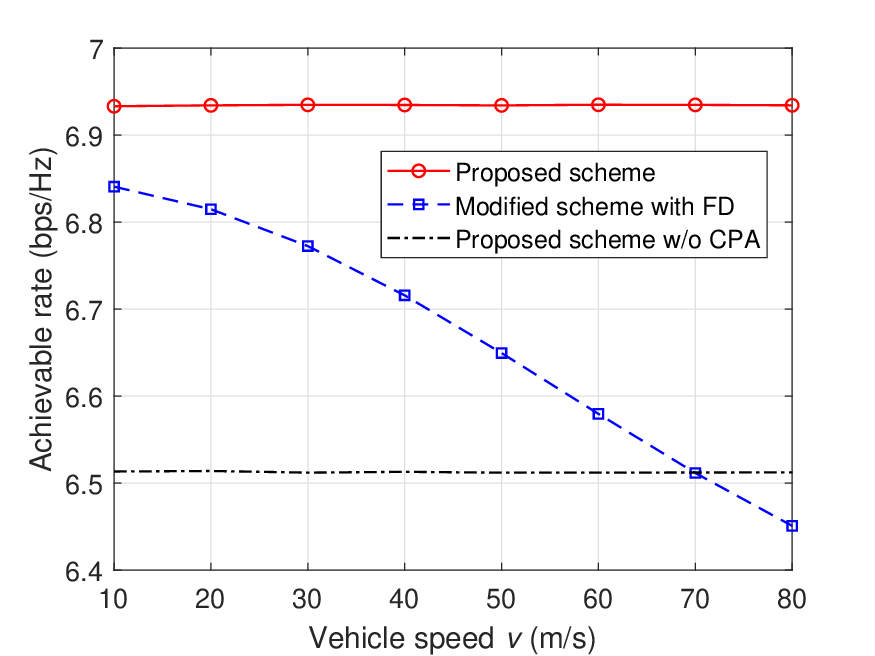}
\caption{Achievable rate $R$ versus vehicle speed $v$.}
\label{rate_speed}
\vspace*{-2em}
\end{figure}

In Fig.~\ref{rate_speed}, we plot the achievable rate $R$ versus the vehicle speed $v$, with $P_t = 26$~dBm, $M = 50$, $\tau_1 = 30$, $T_b = 0.2$~ms, and $K = 10$~dB.
It is observed that by coherently combining the IRS-refracted and non-IRS-refracted channels, the proposed scheme achieves significant rate improvement over that without CPA. Moreover, the achievable rate of the proposed scheme is nearly invariant with respect to $v$ for the case assuming no FD. 
In contrast, the achievable rate of the proposed scheme subject to FD decreases as $v$ increases. This is due to the fact that the channel variation over time is more prominent with higher user mobility, which causes more misalignment of the IRS-refracted and non-IRS-refracted channels for the proposed common refraction phase rotation at the IRS.  

\subsection{Channel Fading Behaviors With versus Without IRS}
\begin{figure} 
     \begin{subfigure}[t]{0.48\textwidth}
         \centering
         \includegraphics[width=\textwidth]{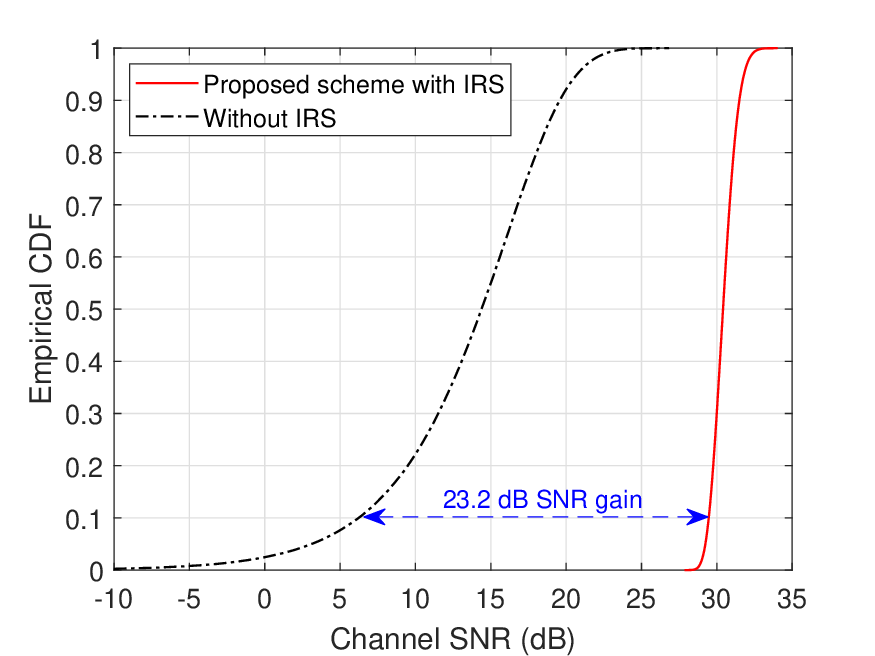}
         \caption{Empirical CDF of the end-to-end channel SNR.}
     \end{subfigure}
     \hfill
     \begin{subfigure}[t]{0.48\textwidth}
         \centering
         \includegraphics[width=\textwidth]{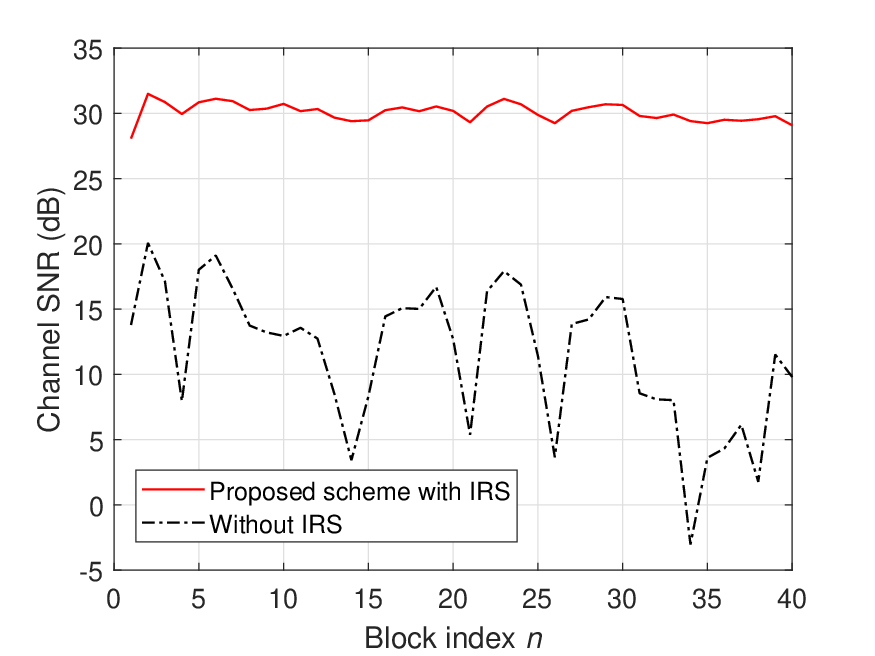}
         \caption{A realization of the end-to-end channel SNRs in one transmission frame.}
     \end{subfigure}
     \caption{Channel fading behaviors of the proposed scheme versus the case without IRS.}
     \label{fts_p}
     \vspace*{-2em}
\end{figure}

Next, we demonstrate in Fig.~\ref{fts_p} the effectiveness of the proposed scheme in converting the BS-user end-to-end  channel from fast to slow fading, with $P_t = 26$~dBm, $M = 50$, $\tau_1 = 30$, and $K = 10$~dB.
We consider the benchmark case without IRS for comparison. 
For our proposed transmission protocol, we define $\gamma^{\left(n\right)} = W^{\left( n \right)}/\sigma^2$ as the effective channel signal-to-noise ratio (SNR) for data transmission at block $n\in \mathcal{N}$, with $W^{\left( 1 \right)} = W_\mathrm{I}$ and $W^{\left( n \right)} = W_\mathrm{II}^{\left( n \right)}$ for $n \in \mathcal{N}_\mathrm{II}$.

In Fig.~\ref{fts_p}(a), we show the empirical cumulative distribution function (CDF) of the effective channel SNR, $\gamma^{\left(n\right)}$, of the proposed scheme as compared to that without IRS, for $n \in \mathcal{N}_\mathrm{II}$. It is observed that owing to the IRS passive beamforming gain in Stage~II, the proposed scheme achieves about 23.2~dB SNR gain over that without IRS at the same outage rate of 10\%. In Fig.~\ref{fts_p}(b), we show one realization of $\gamma^{\left(n\right)}$, for $n\in \mathcal{N}$. It is observed that when $n>1$ (i.e., $n \in \mathcal{N}_\mathrm{II}$), the proposed scheme achieves not only much higher average SNR but also less channel gain fluctuation (i.e., much less fading) in Stage~II, as compared to that without IRS. 

\subsection{Impact of the Number of Refracting Elements $M$}
Moreover, we evaluate the effect of the number of refracting elements $M$ on the achievable rate  performance. We consider the conventional cascaded channel estimation (CCCE) scheme in \cite{ofdmb1} for comparison, where the cascaded channels associated with all the refracting elements and the non-IRS-refracted channel are estimated for each block. In this scheme, the minimum number of pilot symbols required  is $\tau_1 = \tau_2 = M +1$ (with $\tau_\mathrm{D} = 0$ in block 1) for each block and the IRS refraction for data transmission is designed according to  (\ref{optimal_vec}). 
\begin{figure}
\centering
\includegraphics[width=0.60\textwidth]{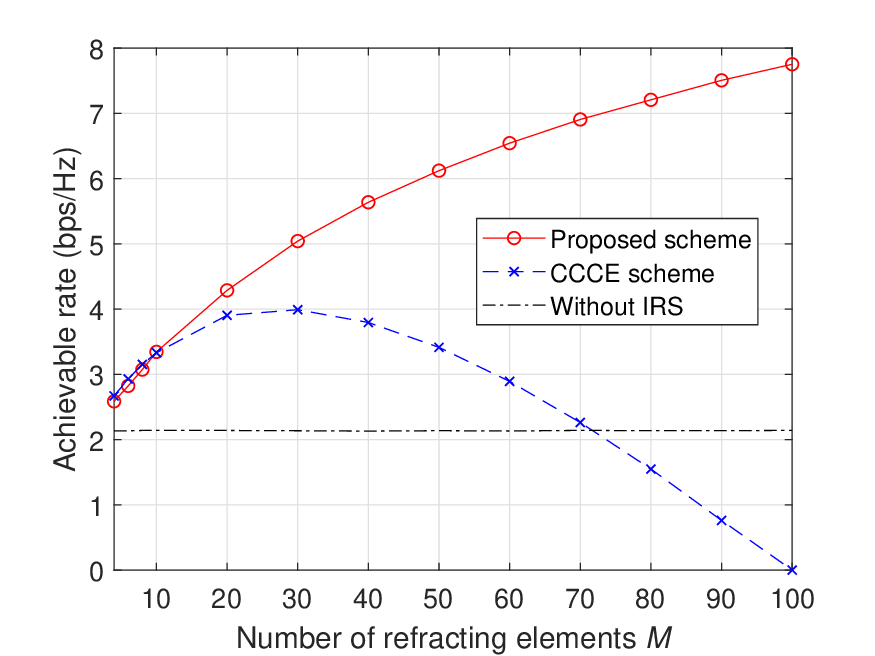}
\caption{Achievable rate $R$ versus $M$.}
\label{rate_mr}
\vspace*{-2em}
\end{figure}
In Fig.~\ref{rate_mr}, we show the achievable rate $R$ versus the number of refracting elements $M$, with $P_t = 26$~dBm,  $\tau_1 = 30$, and $K = 0$~dB. It is observed that the achievable rate of the proposed scheme increases with $M$. This is due to the fact that the proposed scheme efficiently estimates the LoS dominant cascaded channel without the need of increasing the training overhead and also achieves higher passive beamforming gains as $M$ increases. In contrast, the achievable rate of the CCCE scheme first increases and then substantially decreases with $M$, which is due to the increasing training overhead with $M$ for the cascaded channel estimation. Moreover, it is observed that the CCCE scheme performs slightly better than the proposed scheme when $M<10$. This is due to the following two reasons. First, the CCCE scheme estimates the full cascaded CSI, which leads to better beamforming performance. Second, the training overhead of the CCCE scheme is relatively small for a small $M$, which thus will not significantly degrade the transmission rate.

\subsection{Vehicle-Side IRS versus Roadside IRS}
\begin{figure}
\centering
\includegraphics[width=0.40\textwidth]{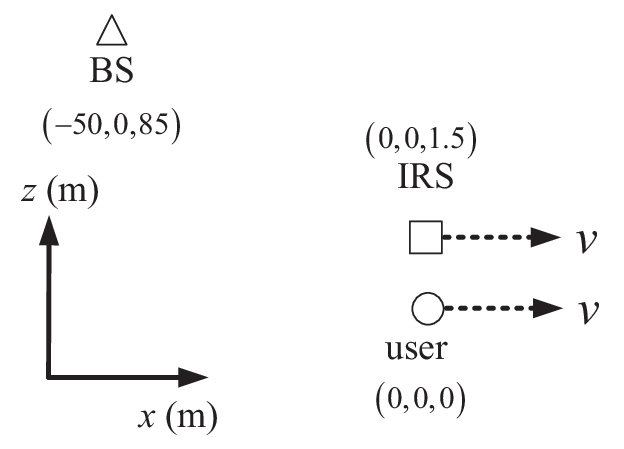}
\caption{Initial node locations in the vehicle-side IRS system.}
\label{init_loca}
\vspace*{-2em}
\end{figure}
\begin{figure} 
\centering
     \begin{subfigure}[t]{0.7\textwidth}
         \centering
         \includegraphics[width=\textwidth]{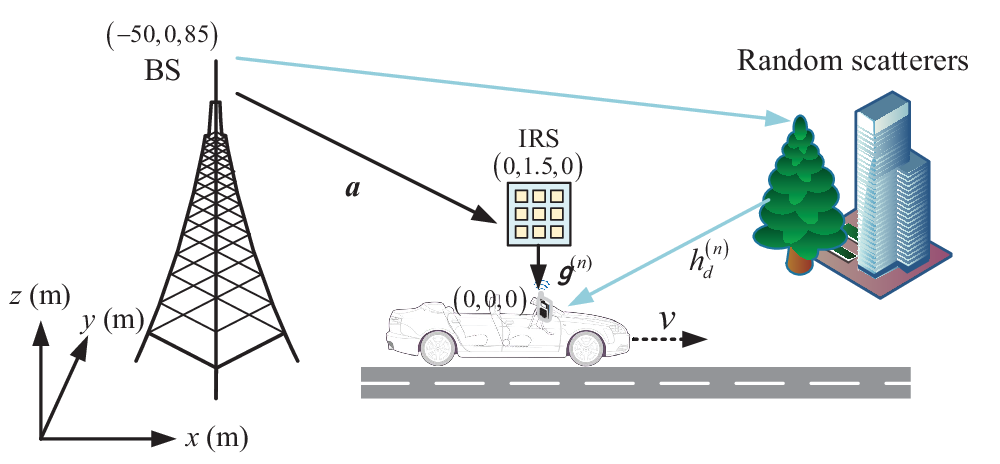}
         \caption{Roadside single-IRS case.}
     \end{subfigure}
     
     \begin{subfigure}[t]{0.7\textwidth}
         \centering
         \includegraphics[width=\textwidth]{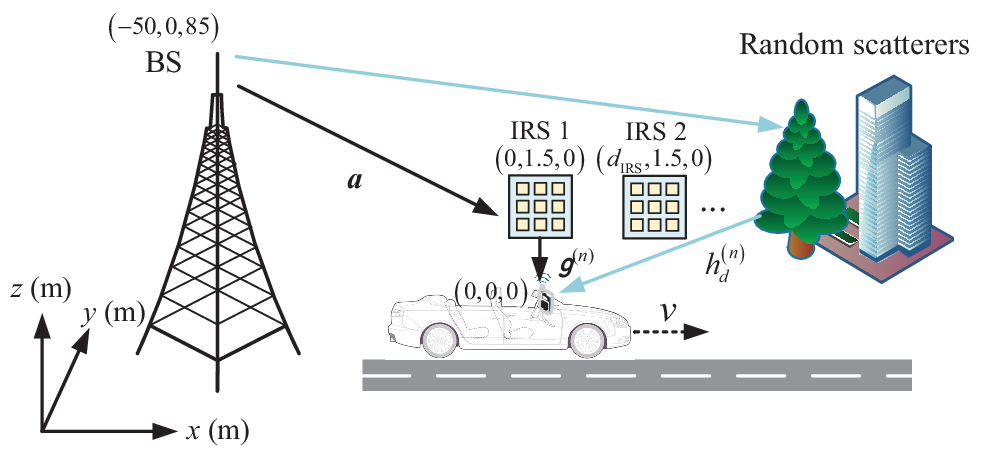}
         \caption{Roadside multi-IRS case.}
     \end{subfigure}
     \caption{Illustration of the roadside IRS-aided communication system for high-speed vehicles.}
     \label{rs_config}
     \vspace*{-2em}
\end{figure}

As shown in Fig. \ref{init_loca}, for the proposed vehicle-side IRS system, the (initial) locations of the BS, IRS, and user are set as $(-50, 0, 85)$~m, $(0, 0, 1.5)$~m, and $(0, 0, 0)$~m, respectively. 
To demonstrate the advantage of our proposed vehicle-side IRS (Intelligent Refracting Surface)  aided high-mobility communication system, in this subsection we consider a roadside IRS (Intelligent Reflecting Surface) assisted high-mobility communication system as the baseline for comparison, which is shown in Fig.~\ref{rs_config}(a).
In this case, instead of installing the IRS on the vehicle, the IRS is deployed at the roadside to assist the communication from the BS to the user, which is equipped with $M$ reflecting (instead of refracting) elements.
For fair comparison, we set the same (initial) locations for the BS and user in both the vehicle-side and roadside IRS deployment cases. 
Moreover, for the roadside IRS case, we set the location of the IRS as $(0, 1.5, 0)$~m.
As such, the minimum IRS-user distance\footnote{Note that in general the IRS-user distance of the roadside IRS case is larger than that of the vehicle-side IRS case, while we set the same distance for the two cases for fair comparison.} (i.e., the shortest IRS-user distance along the user trajectory) is 1.5~m under the roadside IRS case, which is the same as the distance under the vehicle-side IRS case for fair comparison.
Furthermore, the main assumptions for the roadside IRS case are given as follows: i) the BS-IRS channel $\boldsymbol{a}$ is assumed to follow the Rican fading channel model (with $K_\mathrm{RS}$ being the Rician factor) and remain unchanged during the transmission frame due to the fixed locations of BS and IRS;
ii) the IRS-user channel $\boldsymbol{g}^{\left(n\right)}$ is assumed to be LoS with Doppler effect taken into account for modelling the channel variations over time due to the user's high mobility.

\subsubsection{Channel Fading Behaviors}
\begin{figure}
\centering
\includegraphics[width=0.60\textwidth]{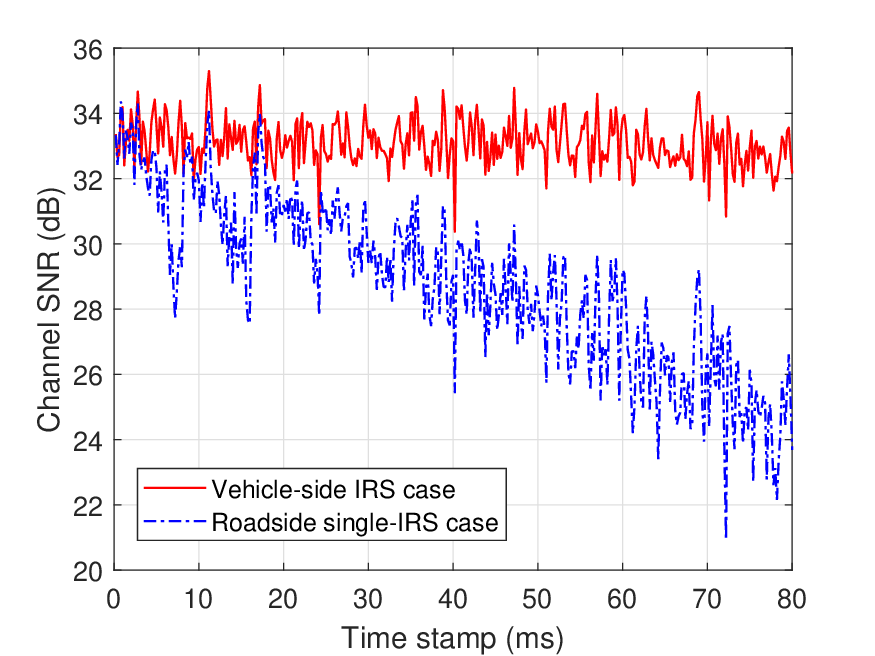}
\caption{Channel fading behaviors of two IRS deployment strategies.}
\label{reali_rv}
\vspace*{-2em}
\end{figure}
In Fig.~\ref{reali_rv}, we plot one realization of the BS-user end-to-end channel SNR during the same traveling period of $T_{tol} = 10 N T_b$ (corresponding to ten transmission frames) for the two IRS deployment cases, with $P_t = 26$~dBm, $\tau_1 = 30$, $T_{tol} = 400 T_b$, $T_b = 0.2$~ms, $M = 50$, and $K = K_\mathrm{RS} = 10$~dB.
It is observed that the proposed vehicle-side IRS deployment case can continuously convert the effective channel from fast to slow fading during the considered traveling period. 
In contrast, for the roadside IRS deployment case with one single IRS at the fixed location (see Fig.~\ref{rs_config}(a)), the corresponding SNR fluctuates dramatically over time and thus can incur higher outage probability. This is because the relative distance between the static IRS and the high-mobility user changes rapidly, which causes high variations in the effective channel phases $\{\psi_x,\psi_y\}$, and also results in severe misalignment of the IRS-refracted channel with $\{\psi_x,\psi_y\}$ estimated at the beginning of each transmission frame.
Moreover, it is observed that the average SNR of the roadside IRS deployment case decreases substantially over time. This is expected since the IRS-user distance under the roadside IRS case increases over time, thus resulting in significantly lower channel gain when the vehicle/user is moving away from the IRS.    

\subsubsection{Achievable Rate Performance}
Due to the limitations of the roadside single-IRS case as discussed above, we further consider the roadside multi-IRS case as shown in Fig. \ref{rs_config}(b).
Specifically, the communication from the BS to the user is consecutively aided by multiple IRSs\footnote{For ease of demonstration, we consider only two IRSs deployed on the roadside, while the results can be easily extended to more than two IRSs for longer transmission time.} deployed at the roadside over constant intervals, each of which is equipped with $M$ reflecting elements.
The inter-IRS distance is set as $d_\mathrm{IRS} = 2$~m. 
During the downlink communication, the nearest IRS to the high-mobility vehicle is selected to serve the user along its trajectory.
Recall that the effective channel phases $\{\psi_x,\psi_y\}$ vary dramatically over time due to the time-varying position of the high-mobility user with respect to the static BS. 
To better track the dynamic variations of $\{\psi_x,\psi_y\}$ so as to achieve a higher passive beamforming gain for data transmission, we set a shorter duration of each transmission frame for the roadside multi-IRS case as $N = 10$, while for both the vehicle-side and roadside single-IRS cases, we set the same frame duration with $N = 40$.

\begin{figure}
\centering
\includegraphics[width=0.60\textwidth]{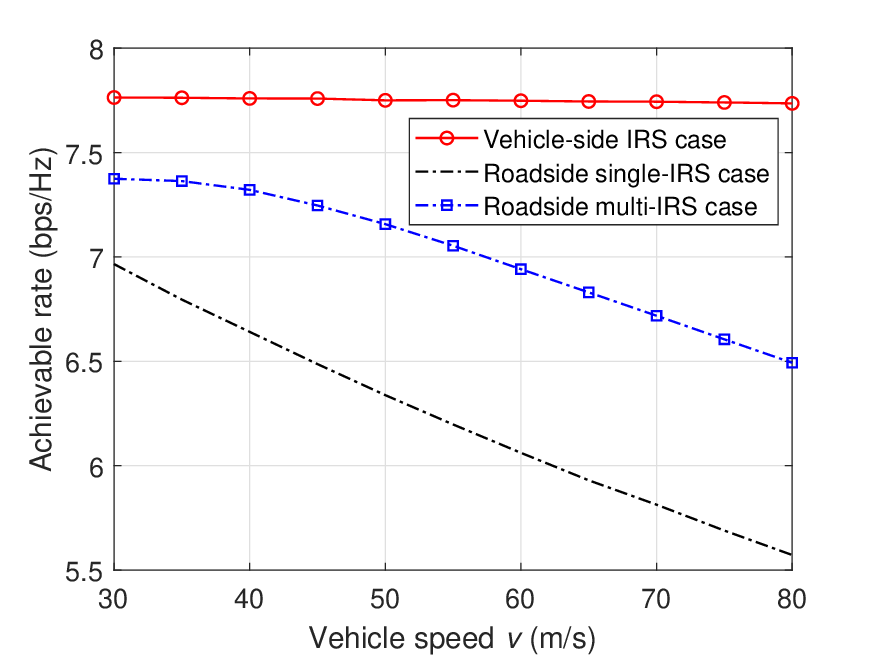}
\caption{Achievable rate performance of three IRS deployment strategies.}
\label{rsvs_rate}
\vspace*{-2em}
\end{figure}

In Fig.~\ref{rsvs_rate}, we plot the average achievable rates versus the vehicle speed $v$ during the same traveling period of $T_{tol} = 400 T_b$ for the three IRS deployment cases, with $P_t = 26$~dBm, $T_b = 0.2$~ms, $\tau_1 = 30$, and $K = K_\mathrm{RS} = 10$~dB. 
It is observed that the roadside multi-IRS case with a shorter frame duration achieves better rate performance than the roadside single-IRS case, especially for higher user mobility, which is due to the following two reasons. 
First, the coverage of the roadside IRS assisted system is extended by employing more IRSs, which helps achieve higher average SNR and less SNR variation over time.
Second, by adopting a shorter duration for each transmission frame, the effective channel phases $\{\psi_x,\psi_y\}$ can be better tracked through more frequent estimation, which alleviates the misalignment problem of the IRS-refracted channel with $\{\psi_x,\psi_y\}$ estimated at the beginning of each frame.
On the other hand, it is observed that the proposed vehicle-side IRS deployment achieves significant rate improvement over both the roadside single/multi-IRS deployment, and the performance gain becomes larger as $v$ increases. This is expected since for the roadside single/multi-IRS deployment, the variations in IRS-user distance and effective phases $\{\psi_x,\psi_y\}$ are more significant as the user mobility increases, thus resulting in lower channel SNR and higher SNR fluctuation, which is in accordance with the observation in Fig.~\ref{reali_rv}. 

\section{Conclusions}
In this paper, we studied a new IRS-aided high-mobility communication system by leveraging the signal refraction function of IRS and employing it with the moving vehicle.
We proposed a new two-stage transmission protocol for achieving efficient channel estimation and IRS refraction design for enhancing the communication rate and reliability.
By exploiting the quasi-static IRS-user channel, the LoS component in the BS-IRS-user cascaded channel is first estimated, based on which the IRS sets its refraction to maximize the passive beamforming gain; while the resultant IRS refracted channel and the non-IRS-refracted channel are estimated subsequently to tune the common phase shift of all IRS refracting elements to align these two channels to further improve the user received signal power for data transmission.   
Simulation results demonstrated that our proposed scheme is effective in converting the end-to-end BS-user channel from fast to slow fading and achieves significant rate improvement over the benchmark scheme designed for slow-fading IRS channels with low-mobility users. 
Moreover, we demonstrated that the proposed vehicle-side IRS (Intelligent Refracting Surface) system is more efficient in improving the user's communication performance in a high-speed vehicle, as compared to the baseline roadside IRS (Intelligent Reflecting Surface) system that requires multiple IRSs deployed with fixed intervals on the roadside to serve the high-mobility users passing by in a consecutive manner. It is also worth noting that there are additional cost and complexity for implementing the road-side IRS due to frequent IRS handover and multi-IRS deployment.

Although this work considered a basic and simplified setup to reveal essential insights for the proposed new vehicle-side IRS system design, its extensions to more general cases such as multi-antenna BS/user, multiple users, frequency selective fading channels, as well as practical discrete phase shift model of IRS, are interesting as well as more challenging to investigate in future work.

\begin{appendices}
\section{}
The search direction of $\{\psi_x,\psi_y \}$ or the gradient of $\frac{\left| \boldsymbol{\xi}^H \left(\psi_x,\psi_y\right)\boldsymbol{y}\right|^2}{\left\| \boldsymbol{\xi}\left(\psi_x,\psi_y\right) \right\|^2}$ is given by
\begin{align}\label{gradient}
\boldsymbol{\Delta }\left(\psi_x,\psi_y\right)&=\nabla \frac{\left| \boldsymbol{\xi}^H \left(\psi_x,\psi_y\right)\boldsymbol{y}\right|^2}{\left\| \boldsymbol{\xi}\left(\psi_x,\psi_y\right) \right\|^2} \nonumber\\
&= \frac{1}{\left\| \boldsymbol{\xi} \right\|^4} \left[\begin{array}{cc}  \left\| \boldsymbol{\xi} \right\|^2\boldsymbol{y}^H\left(\boldsymbol{\xi}_x \boldsymbol{\xi}^H +\boldsymbol{\xi} \boldsymbol{\xi}_x^H\right)\boldsymbol{y} -\left| \boldsymbol{\xi}^H\boldsymbol{y}\right|^2\left(\boldsymbol{\xi}_x^H \boldsymbol{\xi} +\boldsymbol{\xi}^H\boldsymbol{\xi}_x \right)    \\   \left\| \boldsymbol{\xi} \right\|^2\boldsymbol{y}^H\left(\boldsymbol{\xi}_y \boldsymbol{\xi}^H +\boldsymbol{\xi} \boldsymbol{\xi}_y^H\right)\boldsymbol{y} -\left|\boldsymbol{\xi}^H\boldsymbol{y}\right|^2\left(\boldsymbol{\xi}_y^H \boldsymbol{\xi} +\boldsymbol{\xi}^H\boldsymbol{\xi}_y \right) \end{array}\right]
,
\end{align}
where \begin{align}\label{px}
\boldsymbol{\xi}_x = 
\frac{\partial \boldsymbol{\xi}\left(\psi_x,\psi_y\right) }{\partial\psi_x}= \boldsymbol{B}\boldsymbol{V} \left(\left( \boldsymbol{\Xi}_x \boldsymbol{s} \left( \psi_x, M_x \right)\right)  \otimes \boldsymbol{s} \left( \psi_y, M_y \right)\right)
\end{align} 
and 
\begin{align}\label{py}
\boldsymbol{\xi}_y = 
\frac{\partial \boldsymbol{\xi}\left(\psi_x,\psi_y\right) }{\partial\psi_y}= \boldsymbol{B}\boldsymbol{V} \left(  \boldsymbol{s} \left( \psi_x, M_x \right)  \otimes\left( \boldsymbol{\Xi}_y \boldsymbol{s} \left( \psi_y, M_y \right)\right)\right)
\end{align}
denote the partial derivatives of $\boldsymbol{\xi}\left(\psi_x,\psi_y\right)$ with respect to $\psi_x$ and $\psi_y$, respectively, where $\boldsymbol{\Xi}_x = \operatorname{diag}\left(1,j\pi,\ldots,j (M_x-1)\pi\right)\in \mathbb{C}^{M_x \times M_x}$ and $\boldsymbol{\Xi}_y = \operatorname{diag}\left(1, j\pi,\ldots,j (M_y-1)\pi\right)\in \mathbb{C}^{M_y \times M_y}$.
\section{}
For a given random training refraction matrix $\boldsymbol{V}$, the interference-plus-noise term $\boldsymbol{\epsilon}$ in (\ref{collect}) follows the independent and identical complex Gaussian distribution as $\boldsymbol{\epsilon} \sim\mathcal{N}_{c}\left(\boldsymbol{0}, \left( \frac{M\left|\rho\right|^2}{1+K}+\sigma^{2}\right)\boldsymbol{\mathrm{I}}_{\tau_1}\right) $, which can be easily shown via evaluating the Pearson correlation coefficient between any two distinct entries of $\boldsymbol{\epsilon}$ \cite{CORRE}. For notational convenience, define $\boldsymbol{\omega} = \beta^{\left(1\right)} \boldsymbol{V} \boldsymbol{u}\left(\psi_x,\psi_y\right)    + h^{\left(1\right)}_\mathrm{d} \boldsymbol{1}_{\tau_1}$.
As such, the received signal $\boldsymbol{y}$ in (\ref{collect}) is distributed as 
\begin{align}\label{crb_sig_dist}
\boldsymbol{y} \sim\mathcal{N}_{c}\left(\boldsymbol{\omega}, \left( \frac{M\left|\rho\right|^2}{1+K}+\sigma^{2}\right)\boldsymbol{\mathrm{I}}_{\tau_1}\right).
\end{align}
Denote $\boldsymbol{\zeta} = \left[\mathrm{Re}\{\beta^{\left(1\right)}\},\mathrm{Im}\{\beta^{\left(1\right)}\},\mathrm{Re}\{ h^{\left(1\right)}_\mathrm{d}\},\mathrm{Im}\{ h^{\left(1\right)}_\mathrm{d}\},\psi_x,\psi_y\right]^T$ as the collection of relevant parameters. Denote the Fisher information matrix (FIM) by $\boldsymbol{F} \in \mathbb{R}^{6 \times 6}$. Based on (\ref{crb_sig_dist}), each entry of $\boldsymbol{F}$ can be calculated as \cite{estimation_theory}
\begin{align}\label{FIM}
[\boldsymbol{F}]_{\imath,\jmath} = \left(\frac{M\left|\rho\right|^2}{2(1+K)}+\frac{\sigma^{2}}{2}\right)
\mathrm{Re}\{ \left[\frac{\partial \boldsymbol{\omega} }{\partial\zeta_\imath}\right]^H \left[\frac{\partial \boldsymbol{\omega} }{\partial\zeta_\jmath}\right]\} \qquad \imath,\jmath = 1,\ldots,6,
\end{align}
where
\begin{align}
    \frac{\partial \boldsymbol{\omega} }{\partial\zeta_1} &= \boldsymbol{V} \boldsymbol{u}\left(\psi_x,\psi_y\right), & \frac{\partial \boldsymbol{\omega} }{\partial\zeta_2} &= j\boldsymbol{V} \boldsymbol{u}\left(\psi_x,\psi_y\right), \nonumber\\
    \frac{\partial \boldsymbol{\omega} }{\partial\zeta_3} &= \boldsymbol{1}_{\tau_1}, & \frac{\partial \boldsymbol{\omega} }{\partial\zeta_4} &= j\boldsymbol{1}_{\tau_1},\\
    \frac{\partial \boldsymbol{\omega} }{\partial\zeta_5} &= \beta^{\left(1\right)} \boldsymbol{V} \boldsymbol{u}_{x}\left(\psi_x,\psi_y\right), & \frac{\partial \boldsymbol{\omega} }{\partial\zeta_6} &= \beta^{\left(1\right)} \boldsymbol{V} \boldsymbol{u}_{y}\left(\psi_x,\psi_y\right), \nonumber
\end{align}
with $\boldsymbol{u}_{x}\left(\psi_x,\psi_y\right)$ and $\boldsymbol{u}_{y}\left(\psi_x,\psi_y\right)$ being the partial derivatives of $\boldsymbol{u}\left(\psi_x,\psi_y\right)$ with respect to $\psi_x$ and $\psi_y$, respectively. Finally, the CRBs for the effective phases $\{\psi_x,\psi_y\}$ are respectively given by
\begin{align}
    \mathrm{CRB}_x &= \left[ \boldsymbol{F}^{-1}\right]_{5,5}, & \mathrm{CRB}_y &= \left[  \boldsymbol{F}^{-1}\right]_{6,6}.
\end{align}
\end{appendices}

\end{document}